\documentclass[iop]{emulateapj}

\usepackage{graphicx}
\usepackage{amsmath}

\newcommand{\pan}{{\tt Pandurata}\, }
\newcommand{\harm}{{\tt Harm3d}\, }
\newcommand{\pand}{{\tt Pandurata}}
\newcommand{\harmd}{{\tt Harm3d}}

\shorttitle{X-ray Spectra from MHD Simulations}
\shortauthors{Schnittman, Krolik, \& Noble}

\begin{document}

\title{X-ray Spectra from MHD Simulations of Accreting Black Holes}

\author{Jeremy D.\ Schnittman}
\affil{NASA Goddard Space Flight Center \\
Greenbelt, MD 20771}
\email{jeremy.schnittman@nasa.gov}

\author{Julian H.\ Krolik}
\affil{Department of Physics and Astronomy, \\
Johns Hopkins University\\
Baltimore, MD 21218}
\email{jhk@pha.jhu.edu}
\and
\author{Scott C.\ Noble}
\affil{Center for Computational Relativity and Gravitation, \\
Rochester Institute of Technology\\
Rochester, NY 14623}
\email{scn@astro.rit.edu}

\begin{abstract}
We present the results of a new global radiation transport
code coupled to a general relativistic magneto-hydrodynamic simulation of an
accreting, non-rotating black hole. For the first time, we are able to
explain from first principles in a self-consistent way all the components
seen in the X-ray spectra of stellar-mass black holes, including a
thermal peak
and all the features associated with strong hard X-ray emission: a power-law
extending to high energies, a Compton reflection hump, and a broad iron line.
Varying only the mass accretion rate, we are able to reproduce 
a wide range of X-ray
states seen in most galactic
black hole sources.  The temperature in the corona is $T_e \sim 10$ keV in
a boundary layer near the disk and rises smoothly to $T_e \gtrsim 100$
keV in low-density regions far above the disk. Even as the disk's
reflection edge varies from the horizon out to $\approx 6M$ as the
accretion rate decreases, we find that the shape of
the Fe K$\alpha$ line is remarkably constant. This is
because photons emitted from the plunging region are
strongly beamed into the horizon and never reach the observer.
We have also carried out
a basic timing analysis of the spectra and find that the fractional
variability increases with photon energy and viewer inclination angle,
consistent with the coronal hot spot model for X-ray fluctuations. 
\end{abstract}


\section{INTRODUCTION}\label{section:intro}

Since the initial discovery of the magneto-rotational instability
(MRI) by \citet{balbus:91} over two decades ago, tremendous 
progress has been made in simulating astrophysical accretion disks. Large-scale
magneto-hydrodynamic (MHD) simulations have steadily improved their
resolution and physical accuracy, leading to greater understanding of
the fundamental physics of accretion. Yet despite all this success, we
are not much closer to reproducing X-ray observations of accreting
black holes than we were after the original papers on the structure
of their surrounding disks \citep{shakura:73,novikov:73}. From basic
conservation laws, it is relatively easy to understand why the spectrum
seen in some black holes can be ascribed to a multi-color
disk model \citep{mitsuda:84}. However, it was realized very early on
that a great
deal of the power of both stellar-mass black holes and active galactic
nuclei (AGN) is in the form of high-energy X-rays well above the
thermal peak \citep{oda:71,elvis:78}.

Although it is now widely accepted that this hard flux comes from the
inverse-Compton scattering of seed photons from the disk through a hot
corona \citep{liang:78,haardt:93}, we still know little or nothing about
the origin or detailed properties of this corona.  In classical
disk theory\footnote{This is the body of work based on analytic or
semi-analytic models of axisymmetric, usually time-steady, disks whose internal
stresses are assumed to be proportional to the local pressure.} there
is no particular reason even to suppose such a corona exists.  Consequently, almost
all previous work has been perforce phenomenological.  Cartoon sketches
are drawn to suggest the corona geometry, while parameterized models
are used to divide the total dissipation between the disk and corona in an attempt
to fit the data \citep{svensson:94,done:06}. Numerous papers have shown that
for both galactic black holes and AGN the hard spectra commonly observed
can be explained only if the coronal heating is spatially localized and inhomogeneous
\citep{haardt:94,stern:95,zdziarski:96,poutanen:97}, but there
are many
geometries in which this can happen, and most of those proposed are
arbitrary and without firm grounding in dynamics.  At best they
have drawn on qualitative arguments and analogy with the solar
corona \citep{galeev:79}. 
%%JDS
One notable exception is \citet{kawanaka:08}, in which a similar
method to our own is used to couple Monte Carlo radiative transfer to
the global MHD simulations of \citet{kato:04}. However, their simulations are
non-relativistic, assume that the corona is radiatively inefficient, and treat
the corona as dynamically decoupled from the disk proper, so that the
thermal seed photon input is determined without any connection to coronal
properties.

Recent advances in numerical simulation methods now allow us to approach
this problem from an approach founded directly on disk
dynamics.  Angular momentum transport,
the central mechanism of accretion, can be calculated directly, as it
arises from correlations induced by orbital shear in MHD turbulence stirred
by the magneto-rotational instability \citep{balbus:98}.  Moreover, the
same magnetic fields essential to creating internal stresses {\it automatically}
rise buoyantly above the dense regions of the disk and dissipate, creating
a hot corona.  Thus, the mechanisms treated by MHD simulations lead
directly to physical processes that promise to explain coronal phenomenology.

Until now, however, we have lacked the tools necessary for closing the loop
and comparing the results of the simulations directly with the observations
of coronal radiation.  In this paper, by employing the radiative transfer code
\pan \citep{schnittman:12} as a ``post-processor'' to simulation data
generated by the general relativistic MHD code \harm \citep{noble:09},
we make a critical step toward bridging the gap between theory and observation.
In so doing, we will attempt to answer the question, ``Can the coronae predicted
by MHD dynamics produce hard X-rays with the observed luminosity and
spectra?'' 

Our answer to that question will be founded on genuine disk physics.
Whereas analytic disk models rely on dimensional analysis to describe the
scaling of shear stress with pressure, direct calculation of the nonlinear
development of the MRI allows us to compute quantitatively the rate of angular
momentum transfer through the actual magnetic stresses. Thus, we can dispense
with the greatest uncertainties of the traditional accretion model: the dependence of
stress on local physical conditions; the spatial distribution of
dissipation; and the inner boundary condition, which can now be moved
{\it inside} the horizon, and thus made physically irrelevant because
any numerical noise or physical information cannot propagate outward
to the main simulation.  The rate of energy dissipation, rather than being
guessed via some parameterized relation to pressure, can be easily monitored with the
flux-conservative code \harmd, which contains an heuristic cooling
function designed to generate thin disks \citep{noble:09}.  Especially important
to the question we raise about coronal radiation, dissipation in the corona
is the direct result of explicit dynamical calculation, not a scaling
guess about the strength of the magnetic field.  While \citet{shakura:73} and
subsequent phenomenological analyses have been restricted to steady-state,
azimuthally- and vertically-averaged quantities, the MHD 
simulations provide dynamic, three-dimensional information about the
fluid density, 4-velocity, magnetic pressure, gas pressure, and cooling at
every point throughout the computational domain. 

The first step on the path from simulation to observation is to
convert the code variables to physical units by specifying the black
hole mass and accretion rate.  We can then distinguish between the
optically thick disk body and the optically thin corona.
In so doing, we divide the total energy released into a portion
associated with the disk body and a portion associated with the corona.
To predict the spectrum radiated from the disk body,
we make an assumption similar to one made by
\citet{shakura:73} and \citet{novikov:73}, that energy dissipated in
the disk is emitted as thermal radiation from the disk
surface.
To predict the spectrum radiated from the corona, we first demonstrate
that other potentially relevant emission
mechanisms (e.g., thermal bremsstrahlung, synchrotron) are negligible.
We then balance local energy dissipation and inverse-Compton up-scattering
of disk seed photons in order to find the equilibrium electron temperature
and radiation intensity as functions of position throughout the corona.
After both these steps have been accomplished, we use the distribution
of coronal radiation we have just found to predict the hard X-ray
illumination of the disk surface and the Fe~K$\alpha$ fluorescence
line generated by absorption of those hard X-rays.

As described in a companion paper \citep{schnittman:12}, \pan is a
fully relativistic Monte Carlo radiation transport code that
integrates photon trajectories from the disk surface, accounts for scattering
through the hot corona, and transports them to their ultimate destination,
either a distant observer or the black hole horizon.  
For efficient
transport, \pan tracks bundles of many photons along each
geodesic path. These photon packets cover a wide range of energies, further
increasing the efficiency in modeling the broad-band spectra expected
from accreting black holes. While \pan
includes polarization effects in all its scattering calculations, the
results in this paper focus on spectral features alone.

By varying the mass accretion rate, we can reproduce 
%%JDS
many of the features that define
the three main accretion states described in
\citet{remillard:06}: hard, thermal, and steep power law.  Although
the results in this paper are based entirely on
a single simulation whose structure most closely matches the classical
predictions for a disk with $\dot{m}\approx 0.1-0.3$, there is
qualitative agreement with observations spanning the entire range of
$\dot{m} = 0.01-1.0$. 
We have also included a simple model for
fluorescent line production and can reproduce Fe K$\alpha$
features similar to those seen in many galactic black holes and AGN
\citep{miller:04,miller:06a,walton:12}.
For very low accretion rates, the vertically-integrated
optical depth falls below unity in the inner regions, so
for K$\alpha$ production
the disk is effectively truncated around $r \approx 4-6M$. Interestingly, the
iron line profile appears to be independent of the location of the reflection
edge, as long as it is inside the inner-most stable circular orbit
(ISCO).

Lastly, we include some
rudimentary variability analysis, finding results consistent with a large
body of observations: the low-hard state is more variable than the
disk-dominated state, and in both the thermal-dominant and steep power
law states the fractional RMS amplitude increases with photon energy
\citep{cui:99,churazov:01,gierlinski:05,remillard:06}. On short time
scales, the amplitude of 
fluctuations increases with observer inclination angle, consistent
with the coronal hot spot model of X-ray variability.

Although, as we have outlined, numerical simulation data offer many advantages
for spectral predictions, the current state of the art in computational astrophysics
imposes certain limitations.  Two are of particular relevance here.  The
first is that practical and accurate algorithms for treating radiation forces
simultaneously with MHD are still restricted to shearing-box models, and
are not yet ready for application to global models \citep{hirose:06,jiang:12,davis:12}.
As a result, all 3D global simulations to date assume the disk is supported
by a combination of gas and magnetic pressure alone, even though radiation forces
can be an important influence on disk structure, often dominating the disk's
support against the vertical component of gravity \citep{shakura:73}.  
Although \pan provides a fully relativistic radiation
transport {\it post-processor}, its calculation is not incorporated
directly into the \harm simulation, and thus has no effect on the
accretion dynamics. The second limitation is that
because an adequate description of MHD turbulence requires a wide dynamic range
in lengthscales \citep{hawley:11,sorathia:12}, the spatial resolution necessary
to simulate disks as thin as some of those likely to occur in Nature remains beyond
our grasp.  Thus, in some respects, our calculations represent an intermediate
step toward drawing a complete connection between fundamental physics and
output spectra.  Nonetheless, as we will discuss in detail below, they
offer new insights, and, for certain parameter values, are already good enough
approximations to permit direct comparison with observations. 

\section{FROM SIMULATION DATA TO PHOTONS}\label{section:harm}
\subsection{Description of \harm}

The data we analyze for this paper are drawn from the highest
resolution simulation reported in \citet{noble:10} and \citet{noble:11},
designated ``ThinHR'' in those papers.  \harmd, the code used to
generate the data, is an intrinsically conservative 3D MHD code in full general
relativity; this particular simulation was computed in a Schwarzschild
spacetime. Because it uses a coordinate system based on Kerr-Schild, \harm is
able to place the inner boundary of the computation volume inside the
black hole's event horizon, thus obviating the need for any guessed
inner boundary conditions. The stress-energy conservation equation is
modified to include a local cooling function; that is, we write
$\nabla_\nu T^\nu_\mu = - {\cal L}u_\mu$, where $T^\nu_\mu$ is the stress-energy
tensor, $u_\mu$ is the specific 4-momentum, and ${\cal L}$ is
non-zero only for gravitationally-bound gas, and only when the local
temperature is greater than a target temperature
$T_*$.  When the temperature exceeds that threshold, the excess heat is radiated
away on an orbital timescale.  The target temperature $T_*$ is chosen so as to keep
the disk's aspect ratio\footnote{Throughout this paper we will
  consider multiple different scale heights. Here we refer to the
  gas density-weighted scale height $H_{\rm dens}$.} $H_{\rm dens}/r$
close to a single pre-set value at all radii. In dimensionless code units,
$T_* \equiv (\pi/2) (R_z/r)(H_{\rm dens}/r)^2$, where $R_z$ describes the correction to
the vertical gravity due to relativistic effects \citep{noble:10}. For the
ThinHR simulation, the target scale height was $H_{\rm dens}/r=0.06$.

We took special pains to ensure the numerical quality of these simulations.
Every $20M$ in time\footnote{We set $G=c=1$, so time has units of
$(M/M_\odot)\cdot 4.9\times 10^{-6}$ s, and distance has units of
$(M/M_\odot) \cdot 1.5\times 10^5$ cm.}, we measured the number of
cells across the fastest growing MRI wavelength in both the vertical
and the azimuthal directions ($\lambda_z$ and $\lambda_\phi$).  The
minimum number to achieve the correct linear growth rate for vertical
modes is 6 cells per $\lambda_z$ \citep{sano:04}; to describe
nonlinear behavior, at least 20 cells per $\lambda_\phi$ and at least 10
per $\lambda_z$ are necessary \citep{hawley:11}.  The mass- and
time-weighted values in ThinHR were 25 (vertical) and 18 (azimuthal).
As discussed in \cite{hawley:11}, by this and several other measures, ThinHR
is the best-resolved global thin-disk accretion simulation in the literature.
By examining the time-dependent hydrodynamic and radiative properties
of the fluid at several fiducial radii, \citet{noble:10}
determined that the final $5000M$ of the ThinHR
simulation met the relevant criteria for inflow equilibrium in the
inner disk. We therefore restrict our analysis of the simulation data
to that period.

In studying simulations intended to represent statistically steady
accretion, it is important to recognize that when there is only a finite
amount of mass on the grid, some of it must move out in order to absorb
the angular momentum removed from accreted material.  Consequently, the
radial range over which the disk can be said to be in inflow equilibrium
is limited.  For the simulations under consideration here, that range
was typically $r \lesssim 20M$. In \citet{noble:11}, the disk beyond
this radius was simply replaced with a standard relativistic
Novikov-Thorne (N-T) thin disk \citep{novikov:73}.  That paper
focused exclusively on thermal radiation, so a thin disk was an
appropriate extrapolation of the simulation data beyond $20M$. Here, we
are primarily interested in the coronal properties of the accretion
flow, for which there are no simple analytic solutions. Therefore, we
include the entire body of simulation data out to $\sim 60M$, beyond which
the surface density of the gas begins to decrease rapidly, and the
accretion disk is effectively truncated. 

\begin{figure}[ht]
\caption{\label{fig:harm_nt} Luminosity profile $dL/d(\log r)$, integrated
  over $\theta$ and $\phi$, and averaging over time.
  The solid curve is the \harm data, and the dashed curve is the
  Novikov-Thorne prediction. For both cases, the Eddington-normalized
  accretion rate is $\dot{m}=0.1$.}
\begin{center}
\includegraphics[width=0.9\linewidth]{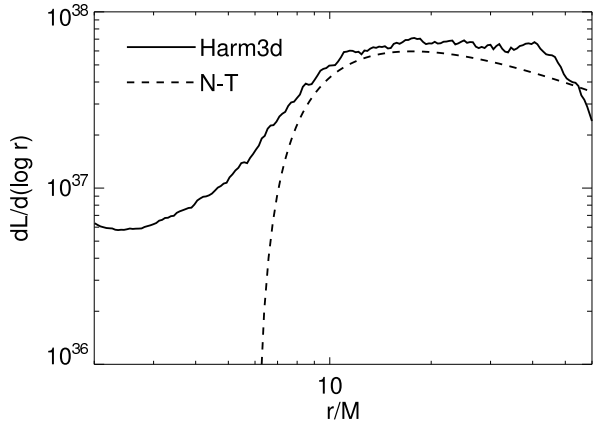}
\end{center}
\end{figure}

Despite the
fact that the disk is not strictly in inflow equilibrium outside of
$\sim 20M$, the dissipation profile still roughly follows that
expected for a classical disk. This can be seen in Figure
\ref{fig:harm_nt}, where we plot the radial shell-integrated
dissipation profile for the \harm data, along with that given by
N-T for the same accretion rate. Unlike \citet{noble:11},
here we make no attempt to normalize the dissipation profile by the
radial mass accretion rate, which explains the somewhat larger
deviation of \harm from N-T outside $\sim 10M$ shown in this figure
than in Figure~2 of \citet{noble:11}. Since the disk surface
temperature scales like $r^{-3/4}$ at large radius, and the corona
temperature also decreases outward (see below, Fig.\ \ref{fig:Te_r}), we
do not expect significant contribution to the X-ray flux from outside
of $r=60M$. 

\subsection{Conversion from Code to Physical Units}
\label{sec:units}

When comparing the \harm predictions with real physical systems, the
first step is to convert the fluid variables from dimensionless
code units to physical cgs units. This conversion requires
specifying the black hole mass $M$, which sets the natural length and time scales,
and the accretion rate $\dot{M}$, which determines the scale for
the gas density, cooling rate, and magnetic pressure.
One technique for enacting this conversion is
described in the appendix of \citet{noble:11}, where the actual
ray-tracing calculation is done in dimensionless units, and only the final
observed spectrum is converted to physical units. That approach works
best for optically thin systems where a single emission mechanism is
used throughout the accretion flow and the photons do not interact
with the matter. 

In this work, however, we are interested in very different radiation
processes in the disk and the corona. Therefore, before we even begin
the \pan ray-tracing calculation, a photospheric boundary must be defined to
distinguish between the cool, dense disk and the relatively hot,
diffuse corona\footnote{Throughout this paper, we use the terms
  ``photosphere'' and ``photospheric boundary''
  interchangably. The term ``corona'' refers to the volume outside of
  the photosphere surface.}. Since the location of this photosphere is a
function of the gas density, it will be different for different values
of the accretion rate.

For this reason, we must convert from code to physical units at the
beginning of the calculation.  There is only one consistent way to
do this.
Following \citet{schnittman:06} and \citet{noble:09}, we relate
the physical density to the code density by
\begin{equation}\label{eqn:code_cgs}
\rho_{\rm cgs} = \rho_{\rm code} \frac{4\pi c^2}{\kappa G M}
\frac{\dot{m}/\eta}{\dot{M}_{\rm code}}\, , 
\end{equation}
where $\kappa=0.4$ cm$^2$/g is the electron scattering opacity, and
$\dot{m}$ is the Eddington-scaled accretion rate assuming a radiative
efficiency $\eta = 0.06$.
For ThinHR, the mean accretion rate in code units was
$\dot{M}_{\rm code} =3\times 10^{-4}$. Regardless of the exact value
for $\eta$ used in the conversion from code units to cgs, \pan
itself results in an independent value for the
radiative efficiency, which is listed in Table \ref{table:L_cor}.
As shown there, it is never far from $\simeq 0.06$ when radiation
from the outer disk is included.

\begin{table}[ht]
\caption{\label{table:L_cor} For a range of mass accretion rates:
the bolometric radiative efficiency $\eta$, the time-averaged fraction of
total luminosity in the corona, the radius of the reflection edge $R_{\rm refl}$,
the disk-corona transition radius $R_{\rm trans}$, and the height $H_{\rm phot}$
of the scattering photosphere (averaged over $r=10-30M$).  The
dependence of $\eta$ on $\dot{m}$ is in
part an artifact of our model, as explained in the text.
Note also that emission outside $R=60M$, ignored here, adds an
additional $\simeq 0.012$
to the radiative efficiency.}
\begin{center}
\begin{tabular}{lccccc}
\hline
\hline
$\dot{m}$ & $\eta$ & $L_{\rm cor}/L_{\rm tot}$ & $R_{\rm refl}/M$ & 
$R_{\rm trans}/M$ & $H_{\rm phot}/r$ \\
\hline
0.01 & 0.056 & 0.40 & 6.1 & 8.8 & 0.11 \\
0.03 & 0.052 & 0.29 & 4.4 & 7.4 & 0.19 \\
0.1  & 0.051 & 0.19 & 2.1 & 6.4 & 0.31 \\
0.3  & 0.048 & 0.13 & 2.0 & 5.7 & 0.43 \\
1.0  & 0.042 & 0.09 & 2.0 & 5.1 & 0.55 \\
\end{tabular}
\end{center}
\end{table}

Once the physical density is specified, the location of the
photosphere at each point in the disk at any particular time is calculated
by integrating the optical depth $d\tau = \kappa\,
\rho(r,\theta,\phi) r\, d\theta$ at constant $(r,\phi)$ from the poles
at $\theta=0,\pi$ down towards the disk. The photosphere is then
defined as the surface where the integrated optical depth reaches
unity. For the top and bottom of the disk, the photospheric surfaces 
can be written as $\Theta_{\rm top}(r,\phi)$ and $\Theta_{\rm
  bot}(r,\phi)$ as in \citet{schnittman:12}:
\begin{equation}\label{eqn:Theta_top}
\int_{\theta=0}^{\theta=\Theta_{\rm top}} d\tau =
\int_{\theta=\Theta_{\rm bot}}^{\theta=\pi} d\tau = 1\, ,
\end{equation}
%%JDS
and the height of the photosphere is then simply given
by $H_{\rm phot} = r|\cos\Theta|$. 
%%

%%JDS
With increasing $\dot{m}$, the photosphere height increases, making the
disk more like a bowl or inverted cone (imagine rotating the contours
of Fig.\ \ref{fig:rho_contour} around the z-axis).  This shape
increases the probability that 
photons scatter off 
other parts of the disk surface (the relativistic version of this
effect is sometimes called ``returning radiation;'' see
\citet{cunningham:76})
and may subsequently be captured by the black hole. Thus, the radiative
efficiency decreases steadily with larger $\dot{m}$.  This effect may
be interpreted as the beginning of ``super-Eddington photon
trapping.''

Just as the gas density must be converted from code units to physical
units, so do the magnetic field and local cooling rate. With
dimensional analysis, determining these conversion factors is
trivial. In cgs units, the magnetic energy density is given by $U_B =
B^2/(8\pi)$, so the conversion factor is simply
\begin{equation}\label{eqn:code_cgs_b2}
\frac{B^2_{\rm cgs}}{B^2_{\rm code}} = c^2\frac{\rho_{\rm
    cgs}}{\rho_{\rm code}}\, .
\end{equation}
The local cooling rate $\mathcal{L}$ has units of energy density per
time, so its conversion factor is given by
\begin{equation}\label{eqn:code_cgs_L}
\frac{\mathcal{L}_{\rm cgs}}{\mathcal{L}_{\rm code}} = c^2\frac{\rho_{\rm
    cgs}}{\rho_{\rm code}} \frac{t_{\rm code}}{t_{\rm cgs}} = 
\frac{c^5}{GM} \frac{\rho_{\rm cgs}}{\rho_{\rm code}} \, .
\end{equation}

\subsection{Disk Structure}

\begin{figure}[ht]
\caption{\label{fig:rho_contour} Fluid density profile for a slice of \harm
  data in the $(r,z)$ plane at simulation time $t=12500M$. Contours
  show surfaces of constant optical depth with $\tau=0.01, 0.1,
  1.0$. Fiducial values for the black hole mass $M=10M_\odot$ 
  and accretion rate $\dot m =0.1$ were used.}
\begin{center}
\includegraphics[width=0.15\linewidth]{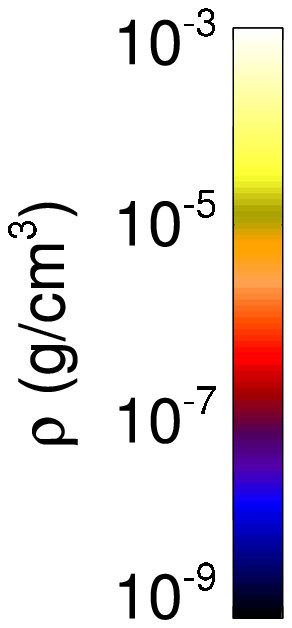}
%\hspace{0.5cm}
\includegraphics[width=0.75\linewidth]{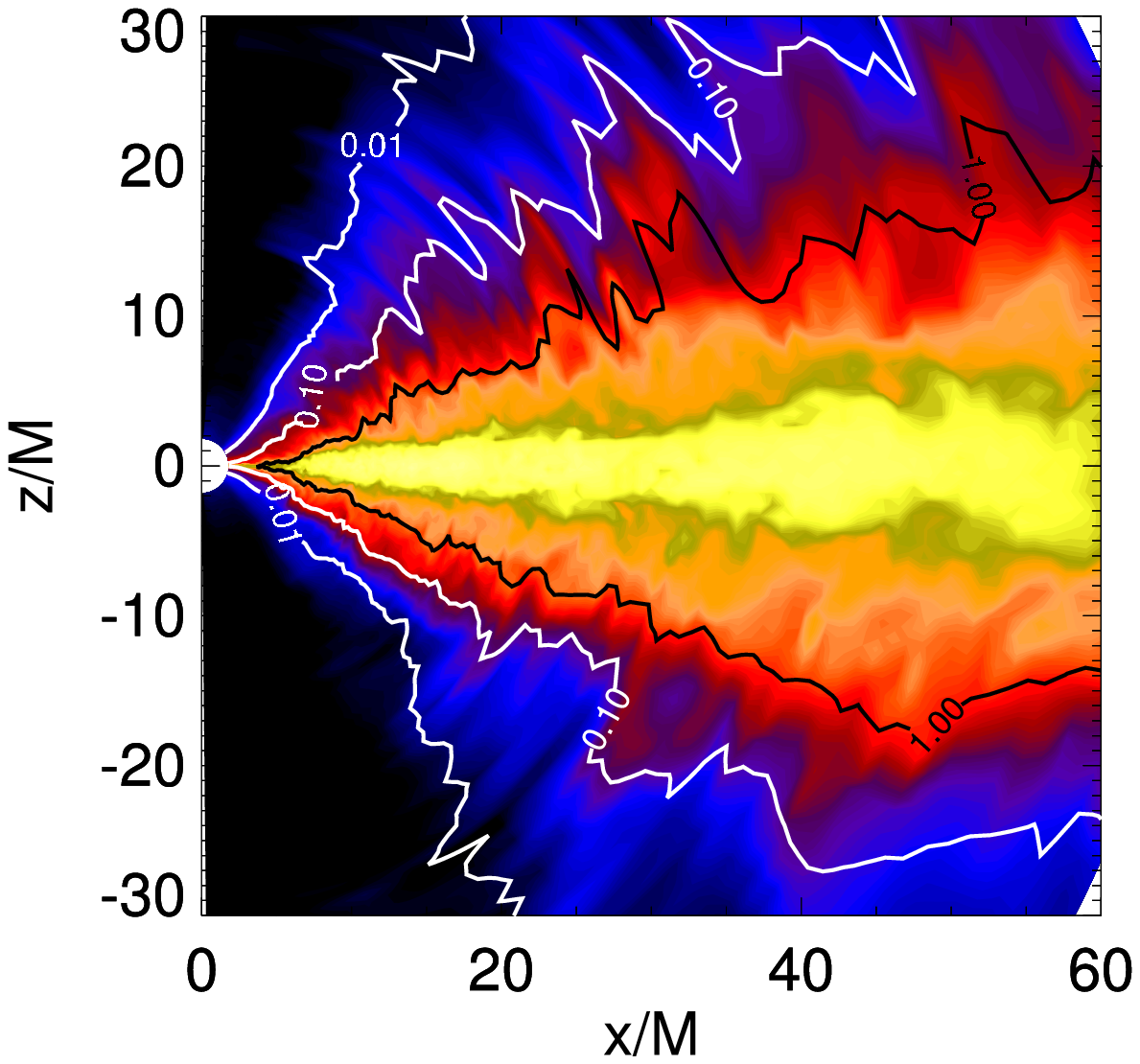}
\end{center}
\end{figure}

Figure \ref{fig:rho_contour} shows a snapshot of the gas density in the
$(r,z=r\cos\theta;\phi=0)$ plane for fiducial values of the black hole mass
$M=10M_\odot$ and accretion rate $\dot{m}=0.1$. The solid contour
lines show surfaces of constant optical depth. Note that while 
the density-weighted scale height of the disk $H_{\rm dens}/r$ is only $\approx
0.06$, the photosphere is located at a height several times that above
the midplane, with $H_{\rm phot}/r \approx 0.3$ in the region of peak
emission $r=10-30M$ for this choice of
accretion rate.  This is to be expected; in stratified shearing box
simulations with careful treatment of thermodynamics and radiation
transfer, the scattering photosphere often lies 3--4 scale heights
from the plane \citep{hirose:09}.

For $\dot{m}=0.1$, the total optical depth of the disk ranges from order
unity in the plunging region up to $\tau \approx 100-200$ in the disk
body at $r>10M$ . Where the total optical depth is less than 2, we say
that there is no disk, only corona (i.e., no solution exists for eqn.\
\ref{eqn:Theta_top}). We denote the radius of this
transition by $R_{\rm refl}$; in the language of \cite{hk:02}, this
is the radius of the ``reflection edge.'' 

\begin{figure}[ht]
\caption{\label{fig:bb_contour} Magnetic energy density profile for
a slice of \harm data in the $(r,z)$ plane corresponding to the same
conditions as in Figure~\ref{fig:rho_contour}.}
\begin{center}
\includegraphics[width=0.15\linewidth]{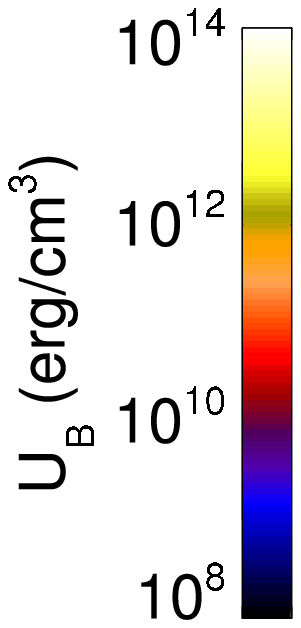}
%\hspace{0.5cm}
\includegraphics[width=0.75\linewidth]{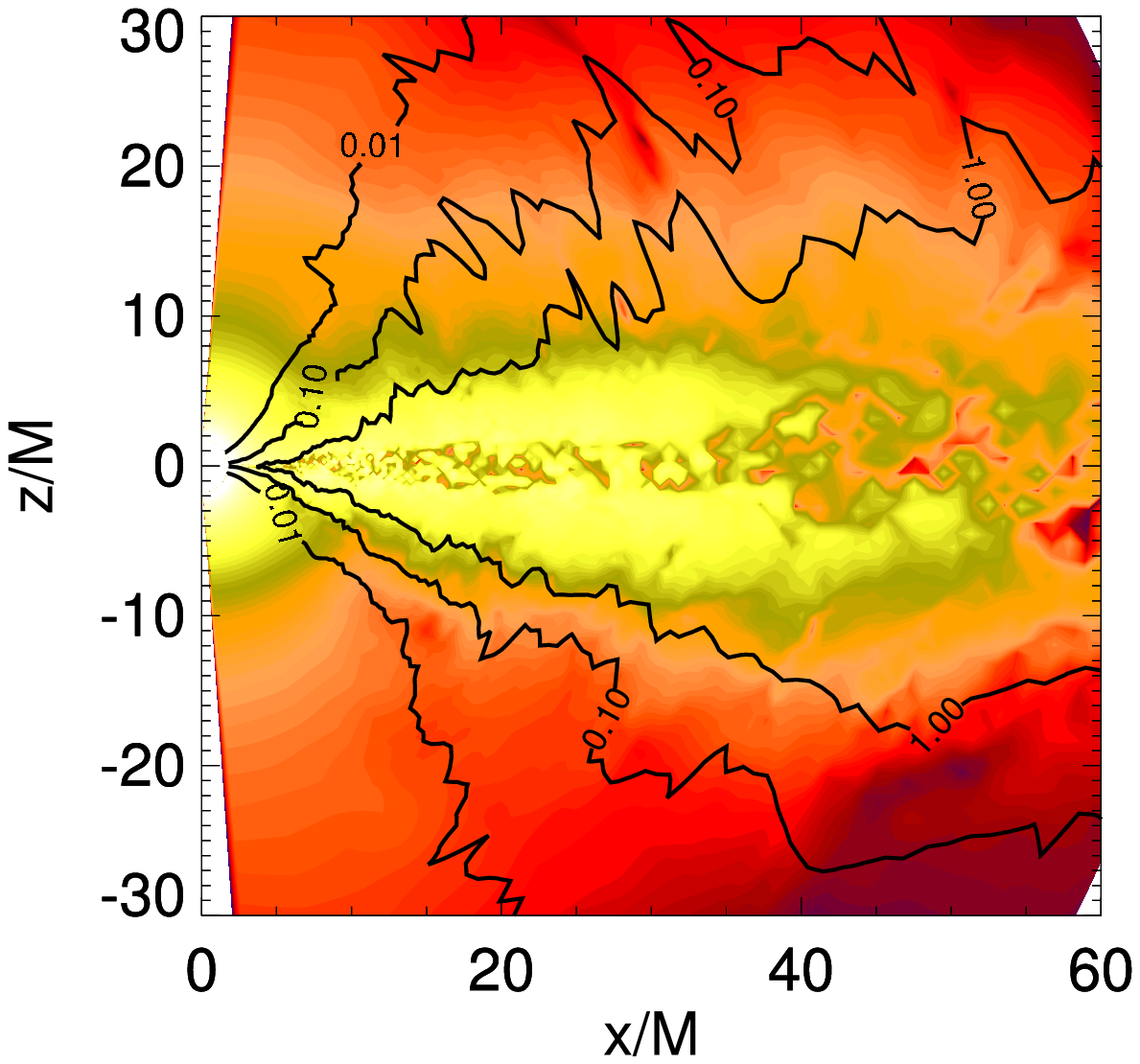}
\end{center}
\end{figure}

\begin{figure}[ht]
\caption{\label{fig:L_contour} Local cooling rate $\mathcal{L}$ for a slice of \harm
data in the $(r,z)$ plane corresponding to the same conditions as
in Figure~\ref{fig:rho_contour}.  Black regions contribute zero emission.} 
\begin{center}
\includegraphics[width=0.15\linewidth]{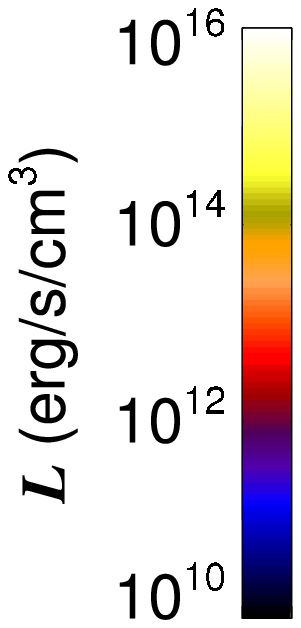}
%\hspace{0.5cm}
\includegraphics[width=0.75\linewidth]{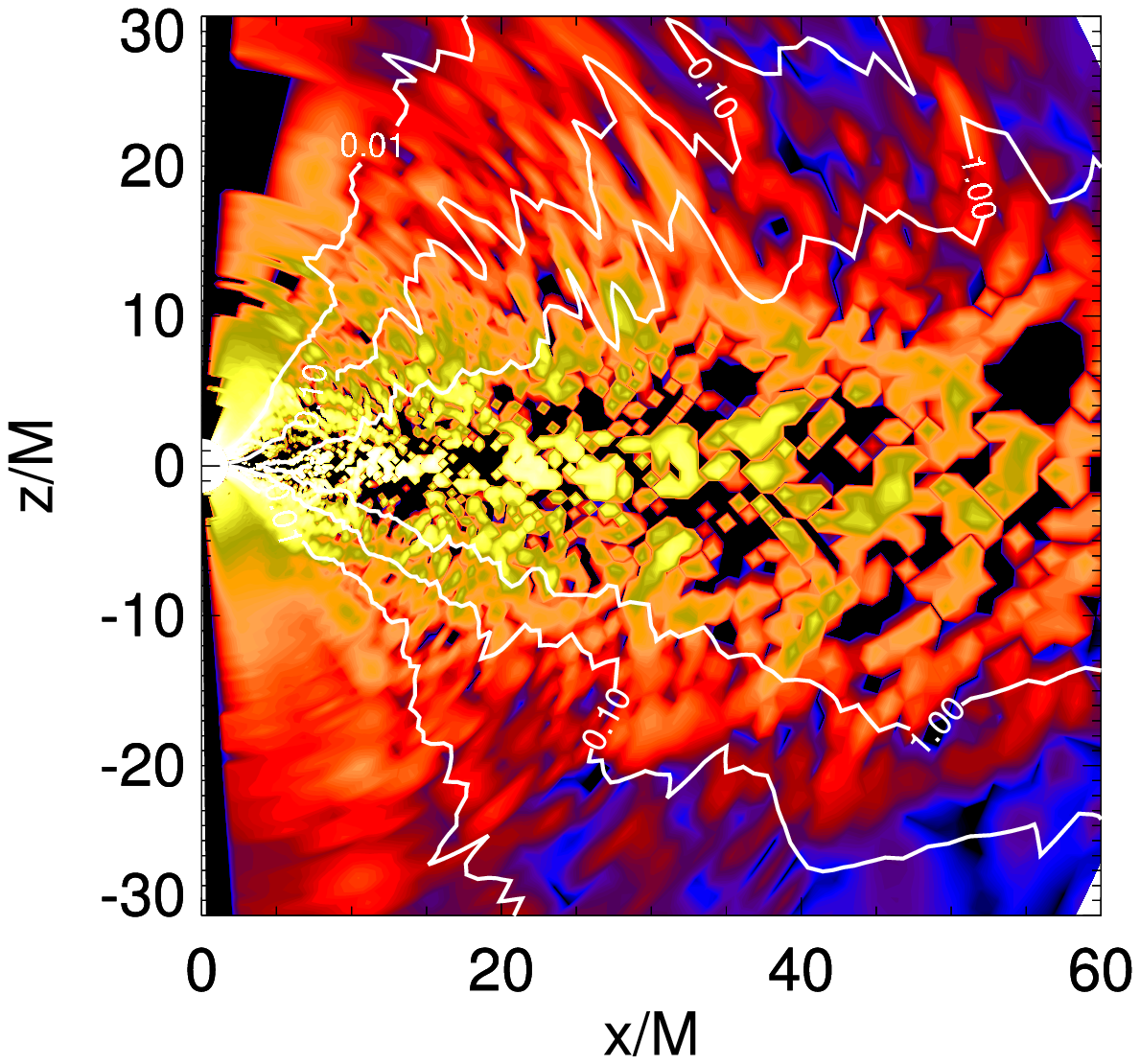}
\end{center}
\end{figure}

In Figures \ref{fig:bb_contour} and \ref{fig:L_contour} we show the
magnetic energy density and local cooling function, respectively. The
\harm data correspond to the same time and the same slice in the $(r,z)$
plane as shown in Figure \ref{fig:rho_contour}, for $M=10M_\odot$ and
$\dot{m}=0.1$. Comparing the gas density and magnetic pressure, we see
that both are concentrated in the disk, but the magnetic scale height
is significantly greater than the gas scale height. This contrast
naturally leads to a corona that is dominated by magnetic pressure, as
seen in most shearing box and global MHD simulations. From equation
(\ref{eqn:code_cgs}), we see that the physical density scales
inversely with black hole mass, but the physical length scale is
proportional to $M$, so the location of the photosphere---and thus the
relative fraction of power from the disk and corona---is
independent of $M$, and depends only on $\dot{m}$. 

\begin{figure}[ht]
\caption{\label{fig:V_L} Fraction of the coronal
  volume that generates a given fraction of the total coronal
  luminosity, for $\dot{m}=0.1$. $10\%$ of the corona is responsible
  for $90\%$ of the emissivity.}
\begin{center}
\includegraphics[width=0.9\linewidth]{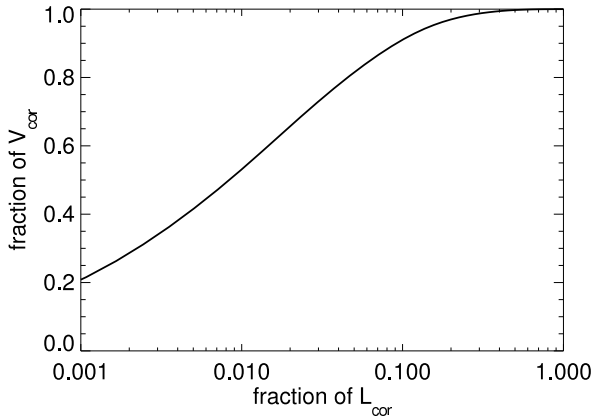}
\end{center}
\end{figure}

The cooling profile appears to closely follow the magnetic field,
consistent with earlier models that use magnetic stress as a proxy for
heat dissipation \citep{beckwith:08}, as well as stratified shearing
box simulations in which the actual dissipation rate is computed
\citep{hirose:06}. As described above, \harm uses a local
cooling function $\mathcal{L}$ to keep the disk relatively thin. This
cooling can also be thought of as the local dissipation of heat, so we
will often identify $\mathcal{L}$ as the emissivity of the gas.  Because
at any given time some of the fluid elements are actually below
their target temperatures, the contours of $\mathcal{L}$ show numerous
isolated patches with no emission (black in Fig.\
\ref{fig:L_contour}). In fact, the coronal emission is
extremely inhomogeneous, and the vast majority of it
comes from a relatively small volume of space. Figure
\ref{fig:V_L} shows the cumulative fraction of the coronal volume
responsible for the cumulative fraction of the coronal luminosity:
$50\%$ of the corona volume generates only $1\%$ of
the luminosity, while $10\%$ of the corona generates $90\%$ of the
luminosity, consistent with many earlier models that assume,
without any dynamical justification,
a highly inhomogeneous heating profile in the corona
\citep{haardt:94,stern:95,zdziarski:96,poutanen:97}.
We now understand that such strongly inhomogeneous
dissipation arises naturally in the MHD simulations, and the simulation
data give a quantitative description of its character.

\begin{figure}[ht]
\caption{\label{fig:cor_frac} Fraction of total dissipation in the
  corona as a function of radius, for a range of accretion rates $\dot{m}$.}
\begin{center}
\includegraphics[width=0.9\linewidth]{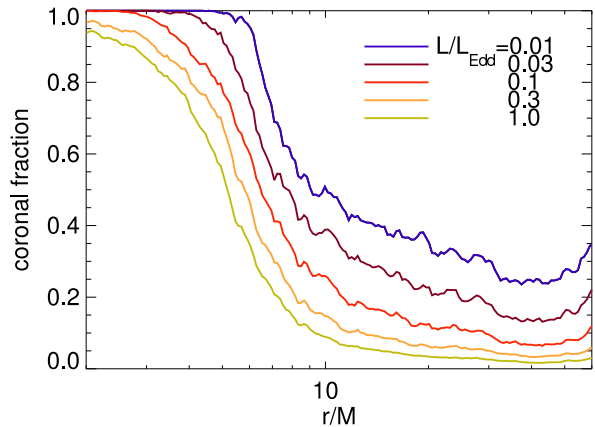}
\end{center}
\end{figure}

As can be seen from Figure \ref{fig:L_contour}, for $\dot{m}=0.1$, the
majority of the total emission comes from within the
disk body, with a sizable contribution from the corona in the
innermost regions. Outside a transition radius $R_{\rm trans}
\approx 6M$, the shell-integrated
luminosity is dominated by the disk, while inside of this radius the
corona dominates. By increasing $\dot{m}$ for the same simulation,
the density scale increases [see eqn.\ (\ref{eqn:code_cgs})],
encompassing a greater fraction of the total luminosity within the
optically thick disk. Conversely, for small values of $\dot{m}$, the disk shrinks
and the corona becomes more dominant. 
Table \ref{table:L_cor} shows how the relative contributions of the
disk and corona change with $\dot{m}$, as well as the locations of
$R_{\rm refl}$ and $R_{\rm trans}$.
The corona dominates throughout the plunging region for all sub-Eddington
values of $\dot m$. This can also be seen in Figure \ref{fig:cor_frac},
where we have plotted the fraction of total dissipation (angle- and
time-averaged) occurring in the corona at each radius for a range of
$\dot{m}$. This fraction can be directly compared to the $f$ parameter
used in the coupled disk-corona model of \citet{done:06}, where
the total dissipation at each radius is divided between disk and
corona.  Thus it should come as no surprise that
our resulting spectra (see below in Sec.\ \ref{section:broad_band})
are qualitatively similar to those that they predicted for comparable
values of $f$. 

\subsection{Global Radiation--Matter Thermal Balance}

All the radiation processes we treat are thermal at some level.  However,
defining the temperature in this context involves some subtlety, both
inside and outside the disk.  Given the density, internal energy, and an
equation of state, it is possible to derive a gas temperature directly from the simulation
data.  However, because \harm does not include any radiation pressure
in its equation of state, this inferred temperature would not be very meaningful.
Inside the optically thick disk, the code temperature should be regarded
as a measure of the total thermal (i.e., gas plus radiation) pressure per unit mass
supporting the gas against the vertical component of gravity, but this is not necessarily
the temperature relevant for its radiated spectrum.  The reason is that,
for much of the interesting range of accretion rates and radii, the material's
support is in fact dominated by radiation pressure.  Consequently, the actual
gas temperature is considerably lower than the code temperature.  In the corona,
although radiation forces should be less important, the code temperature is once
again unsuitable for spectral considerations because it is tied to the
{\it ad hoc} target temperature used by the simulation in order to regulate
the disk's geometric thickness.  Fortunately, though, it is possible to
use other simulation data to self-consistently determine the coronal
temperature through consideration of explicit thermal balance.  Moreover,
unless the thermal balance temperature turns out to be considerably higher
than the code temperature, the dynamics of the simulation remain self-consistent
because support against gravity in the corona is primarily magnetic:
the plasma $\beta$ is in the range $\sim 0.03$--0.3 almost everywhere in the corona.
Thus, the gas temperature (whether code or Compton equilibrium)
has little influence on the density structure in
the corona.  In the remainder of this subsection we describe
how we estimate the genuine physical temperature both inside and
outside the disk photosphere.

Inside the disk (i.e., between the two electron scattering
photospheres), we assume all the emitted radiation is able to
thermalize, and all the heat generated within the disk is
radiated from the same $(r,\phi)$ where it is made.  These
two assumptions allow us to define an
effective temperature at each point on the disk surface:
\begin{equation}
\sigma T_{\rm eff}^4(r,\phi) = \frac{1}{2}\int_{\Theta_{\rm
    top}}^{\Theta_{\rm bot}} \mathcal{L}(r,\theta,\phi)dl\, ,
\end{equation}
where $\sigma$ is the Stefan-Boltzmann constant and the factor of
$1/2$ is due to the fact that half the radiation is emitted from the
top and half from the bottom of the disk. 

In the regions of the disk where most of the flux is generated,
electron scattering opacity is always much greater than the opacity
due to other processes such as free-free absorption \citep{shakura:73}.
We therefore assume the mean intensity in the fluid frame at the photosphere
has the spectrum of a diluted black-body:
\begin{equation}
J_\nu = f_{\rm hard}^{-4}B_\nu(f_{\rm hard} T_{\rm eff})\, ,
\end{equation}
where $B_\nu(T)$ is the Planckian black-body function, and the
hardening parameter $f_{\rm hard}$ is taken to be 1.8
\citep{shimura:95}. As described in \citet{schnittman:12}, the disk
flux also has an angularly dependent intensity
(limb-darkening) and polarization \citep{chandra:60}. 

In the corona, the picture is not so simple. Unlike in the disk, in the
corona the radiation is not expected to be
thermalized. We have considered the contributions from a
number of different radiation mechanisms, including bremsstrahlung,
cyclo/synchrotron, and inverse Compton (IC), but, at
least for the stellar-mass black holes of interest here, we find the coronal power is
completely dominated by inverse Compton. 

Bremsstrahlung and synchrotron emission are both fundamentally local processes,
and depend only on the local electron density, temperature, and
magnetic field. Since the density, magnetic field, and net
emissivity are given by the \harm data, to solve for the temperature
we could simply invert the following equation at each point in the
corona if they were the only cooling agents:
\begin{equation}
\mathcal{L}=P_{\rm brem}(\rho,T_e)+P_{\rm synch}(\rho,T_e,B)\, ,
\end{equation}
where $P_{\rm brem}$ and $P_{\rm synch}$ are the local bremsstrahlung
and synchrotron power density (in erg/s/cm$^3$), respectively. But
this approach is incomplete because the corresponding absorptive
opacity
\begin{equation}
\alpha_\nu = \frac{j_\nu}{B_\nu(T_e)}
\end{equation}
can also be important.  Here $\alpha_\nu$ and $j_\nu$ are the absorption and emission
coefficients respectively for either bremsstrahlung or synchrotron.
For typical coronal conditions of $T_e \sim$ 10--1000 keV, $n_e \sim
10^{16}$--$10^{18}$ cm$^{-3}$, and $B \sim 10^6$--$10^7$ G, we find
that free-free emission and absorption are both negligible, while
synchrotron emission can actually contribute a significant fraction of
the total cooling function. However, the typical cyclotron frequency
for these parameters lies in the infrared, where self-absorption
is strong. Since the corona is optically thick to synchrotron
radiation, it does not end up contributing significantly to the total
cooling: every photon that is emitted is almost instantly re-absorbed.
We are thus left with IC as the dominant emission process in the
corona. 

Unlike bremsstrahlung or synchrotron, IC is a fundamentally {\it non-local}
process because it requires a population of seed photons to be up-scattered
by the hot electrons.  Moreover, the IC seeds can come from distant parts of
the accretion disk.  Local treatments are therefore insufficient.

For a mono-energetic population of electrons with isotropic velocity $v$,
the IC power is \citep{rybicki:04}
\begin{equation}\label{eqn:P_compt}
P_{\rm IC} = \frac{4}{3}\sigma_T c \gamma^2 \beta^2 n_e U_{\rm ph}\, .
\end{equation}
Here $\sigma_T$ is the Thomson cross section, $\beta=v/c$,
$\gamma = (1-\beta^2)^{-1/2}$, $n_e$ is the electron density, and
$U_{\rm ph}$ is the energy density
of the local photon distribution. This local photon density is
not known {\it a priori} from the simulation data, so it must be solved
for using radiation transport.
Because the corona has an optical depth of order unity (i.e., neither
optically thin nor optically thick approximations can be used) and
its geometry is complex, we use Monte Carlo ray-tracing including scattering
to model the transport in a global manner \citep{schnittman:12}.

As described in \citet{schnittman:12}, at every step along its geodesic
trajectory a photon packet has a small chance that it will scatter off an electron. 
When a scattering event occurs, the electron velocity is selected by choosing
a particular velocity in the local fluid frame from the relativistic
Maxwell-Boltzmann distribution for the local temperature. The
direction of the scattered photon packet is chosen from the
probability distribution associated with the Compton scattering
partial cross section. The change in energy of the scattered photons is then
calculated exactly in terms of relativistic kinematics.  When a photon
packet encounters the disk photosphere, its new direction is chosen
from the probability distribution given by \cite{chandra:60}.  At the
temperatures characteristic of the inner regions of accretion disks in
black hole binaries, few medium-$Z$ elements are unstripped, so the
reflection albedo should be high (we treat Fe~K photoionization separately;
see Section~\ref{section:iron_lines}).  Nonetheless, the packet does
lose energy by conventional Compton recoil because the electrons in
the disk surface are ``cold,'' i.e., $kT_e \ll m_ec^2$.

A somewhat similar method has been used by \cite{kawashima:12} to
predict the spectrum from black holes accreting at super-Eddington
rates.  It resembles ours in that they employ Monte Carlo ray-tracing
including IC scattering to predict X-ray spectra.  It differs
from ours in that the underlying data is drawn from an axisymmetric
MHD simulation; this restrictive symmetry assumption prevents
the simulation from properly following the development of MHD
turbulence.  The simulation is pseudo-Newtonian rather than
relativistic, and the ray-tracing likewise assumes a flat spacetime.
In addition, their coronal electron temperature is
not made self-consistent with the simulation's dissipation rate.  They
assume that $T_e$ is the same as the code temperature, but the code
temperature is computed by balancing dissipation with a thermal
equilibrium cooling rate (i.e., it is $\kappa_{\rm abs}B$, where
$\kappa_{\rm abs}$ is free-free opacity and $B$ is a frequency-integrated
Planck function) rather than with the relevant mechanism, inverse
Compton scattering.
A still cruder version has been employed by \cite{you:12}, in which
Monte Carlo radiation transfer with electron scattering in a relativistic
background was also used to predict an output IC spectrum,
but with guessed electron densities, a guessed fraction of the disk
dissipation put into the corona, and a coronal velocity simply set
equal to the local orbital velocity.

With the assumption that the photon diffusion time is short compared to
the time required for any dynamical or thermal changes, the problem can be
thought of as a boundary value problem: given the fluid
density, 4-velocity, and cooling rate at every point in the
corona, along with the thermal seed photon distribution at the
photosphere surface, we need to solve for the electron temperature
$T_e(r,\theta,\phi)$ and photon energy density $U_{\rm ph}(r,\theta,\phi)$
at every point in the corona. To do so, we employ an iterative technique as
follows: 
\begin{itemize}
\item Initially estimate the local value of the radiation density in terms of
 the thermal contribution at the surface of the disk: $U_{\rm
 ph}(r,\theta,\phi)=c\sigma T^4_{\rm disk}(r,\phi)$.
\item Solve equation (\ref{eqn:P_compt}) with $P_{\rm IC}=\mathcal{L}$
  to get $\gamma(r,\theta,\phi)$ throughout the corona. Derive the
  electron temperature at each point from the
  relation $T_e = \frac{2}{3}\frac{m_e c^2}{k_B}(\gamma-1)$ (the
  non-relativistic expression works well as an initial guess for the
  corona temperature).
\item Carry out a complete Monte Carlo ray-tracing calculation with \pand,
  using thermal seed photons from the disk photosphere propagating
  through the corona via Compton scattering.
\item For each volume element in the corona, determine
  the total amount of IC power generated in that zone by
  comparing the ingoing and outgoing energy of every photon packet that
  scatters within that zone.
\item Compare the coronal power from \pan with the cooling
  function $\mathcal{L}(r,\theta,\phi)$ given by \harmd.
  Where the coronal power from the ray-tracing calculation exceeds
  the cooling rate in the simulation, the initial guess for
  $U_{\rm ph}$ was too low, giving a $T_e$ that is too high, and vice
  versa. 
\item Revise the coronal temperature estimates up or down accordingly,
  and repeat the full ray-tracing calculation, getting a new 3D map
  of the cooling function, which is again compared with the target values
  from the simulation data.
\item Repeat until the global solution for $T_e$, $U_{\rm ph}$, and
  $\mathcal{L}$ is self-consistent throughout the corona.
\end{itemize}
%%JDS
This iterative procedure is similar to the technique briefly described
in \citet{kawanaka:08}, where they attempt to balance electron cooling
from IC with electron heating
from Coulomb collisions with energetic ions.

Because the Monte Carlo technique is inherently noisy, the \pan
calculation and the \harm target cooling rate for any individual fluid
element are unlikely to agree very well. We typically use $\sim 10^{7-8}$
photon packets for a single snapshot, while the simulation volume
contains roughly $10^6$ cells.  Consequently, for the majority of the
corona volume, where $\tau < 0.1$, a given cell might see an average
of only {\it one} photon packet.  It should therefore not be surprising that
point-to-point Poisson fluctuations are quite large. Furthermore, as
can be seen in Figure \ref{fig:L_contour}, even the target cooling
function is highly non-uniform, characterized by large-amplitude
fluctuations on a small spatial scale. 

For this reason, before comparing the results of the ray-tracing
calculation with the target \harm emissivity, we apply a smoothing
kernel to both data sets to remove the fluctuations described
above. This smoothing is not only numerically helpful, but is also
strongly motivated from a physical point of view. For any radiation
transport problem in a roughly steady state, the photon energy and
momentum density cannot change significantly over length scales much
shorter than the mean free path. Thus, when testing for convergence of
$U_{\rm ph}$, it is eminently reasonable to
smooth the cooling function $\mathcal{L}$ over the characteristic
scattering length.

In fact, by smoothing over an even greater volume, we can
significantly improve the efficiency of our iterative solver. For
example, if the smoothing length is comparable to the coronal scale
height, then instead of trying to sample $\mathcal{L}$ in $10^6$ fluid
elements, we are effectively only probing 10--100, and thus can use
many fewer photon packets.
After converging at low resolution, we
repeat the calculation with more photons and progressively shorter smoothing
lengths until \pan and \harm agree to high accuracy everywhere down
to the gridscale.

\begin{figure}[ht]
\caption{\label{fig:dL_dr} Instantaneous luminosity profile $dL/d(\log
  r)$, integrated
  over $\theta$ and $\phi$, considering
  only coronal emission. The black curve is the \harm data, and the
  red curve is the \pan ray-traced reconstruction. The
  luminosity is $0.1 L_{\rm Edd}$. The Monte Carlo calculation used $\simeq
  5\times 10^7$ photon packets for this snapshot.}
\begin{center}
\includegraphics[width=0.9\linewidth]{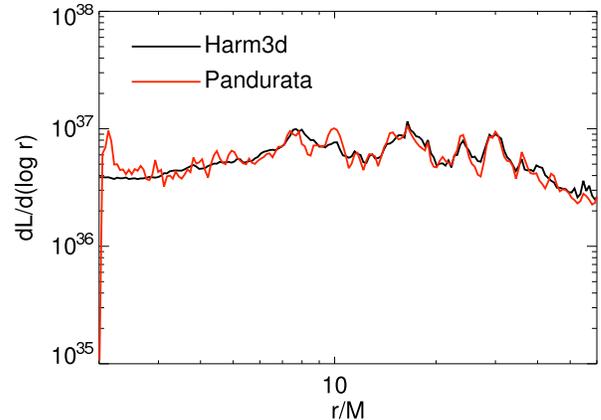}
\end{center}
\end{figure}

\begin{figure}[ht]
\caption{\label{fig:dL_dth} Coronal luminosity profile as in
  Fig.\ \ref{fig:dL_dr}, but for $dL/d(\cos\theta)$.}
\begin{center}
\includegraphics[width=0.9\linewidth]{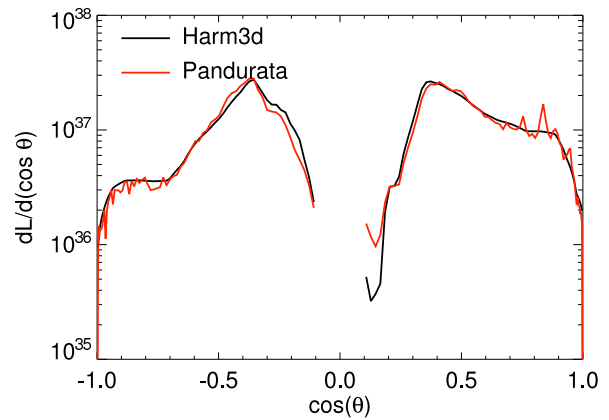}
\end{center}
\end{figure}

One way to see this agreement at a quantitative level is to compare
the radial distribution of coronal emission $dL/d(\log r)$ as derived from
the two codes for a single snapshot, shown in Figure \ref{fig:dL_dr}.
After only 3 levels of iteration, we
are clearly able to resolve coronal hot spots as small as $\Delta r/r \sim
0.2$. By plotting $dL/d(\cos\theta)$, we see that the vertical profile of
the corona is also well-matched (see Fig.\ \ref{fig:dL_dth}).

\subsection{Coronal Temperature}\label{section:cor_temp}

\begin{figure}[ht]
\caption{\label{fig:Te_contour} Electron temperature in the corona for a
  converged solution of the global radiation field, for the same
  snapshot as in Figure \ref{fig:rho_contour}. Within the disk
  photosphere, all the radiation is thermalized, and we assume the
  temperature is uniform for constant $(r,\phi)$.}
\begin{center}
\includegraphics[width=0.15\linewidth]{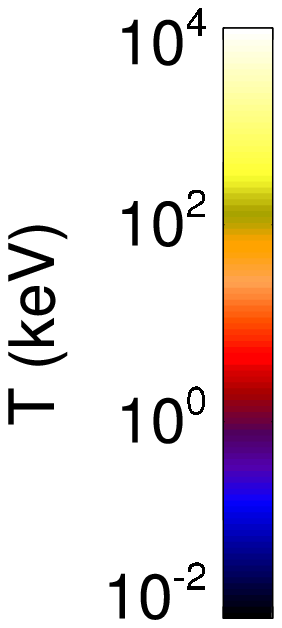}
%\hspace{0.5cm}
\includegraphics[width=0.75\linewidth]{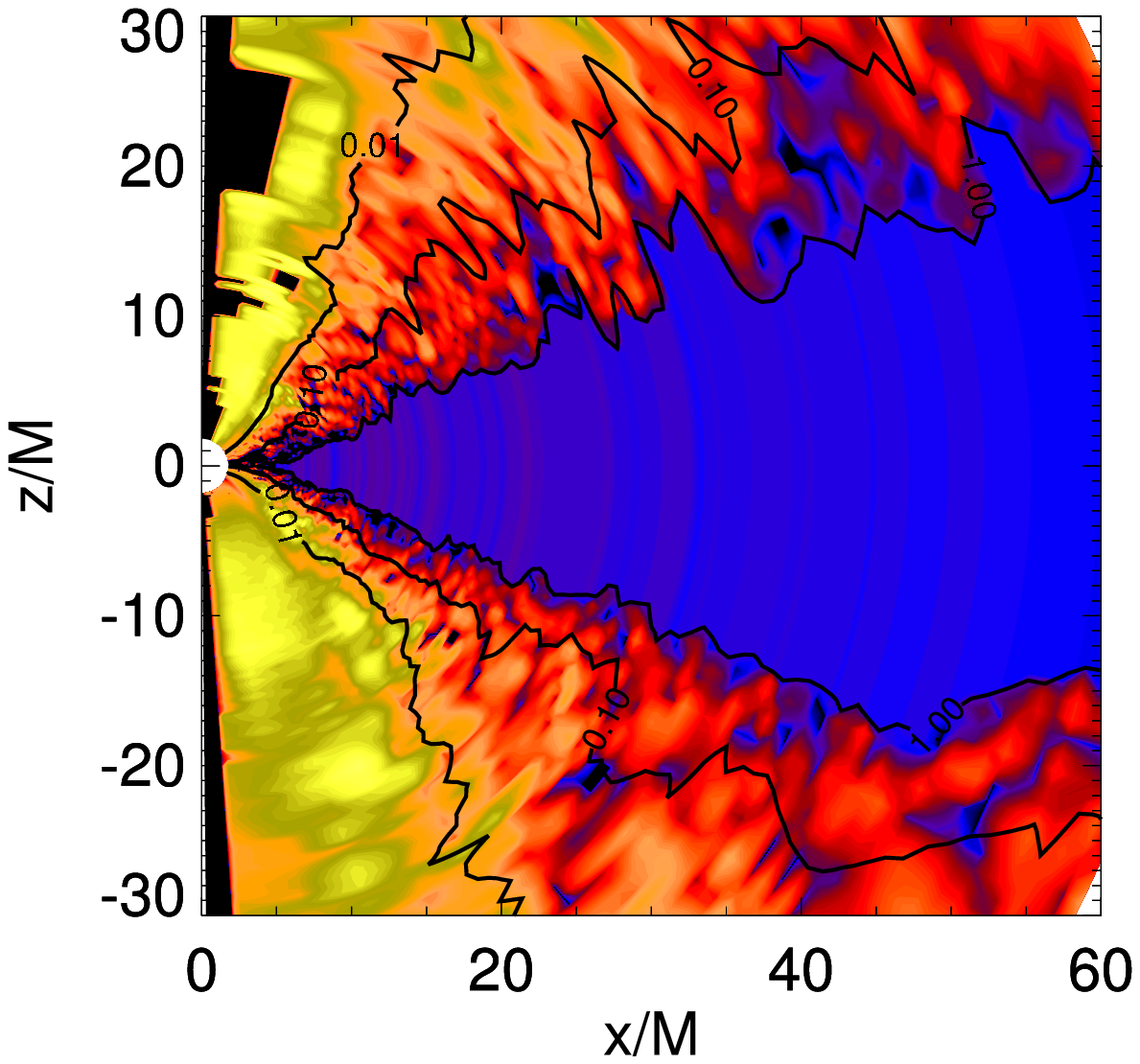}
\end{center}
\end{figure}

The global solution for the electron temperature is shown in Figure
\ref{fig:Te_contour}. The temperature within each vertical slice of
the disk body is constant, and of the order 0.2--1 keV for these
parameters. The corona is much hotter, with $T_e$ ranging
from $\sim 10$--100 keV for $\tau$ between 0.01 and 1. By comparing
Figures \ref{fig:rho_contour} and \ref{fig:L_contour}, we see that the
electron density falls off faster with increasing altitude from the
disk than the dissipation ($H_{\rm diss}\simeq 3H_{\rm dens}$),
leading to higher coronal temperatures as more power
must be released by a smaller quantity of gas. The temperature map also
shows large fluctuations over small spatial scales, yet not quite as
large as those seen in 
$\mathcal{L}$. This is because the regions of high dissipation are
correlated with regions of high density, which has the effect of
smoothing out the temperature gradients [see eqn.\ (\ref{eqn:P_compt})]. 

By changing the $\dot m$ used in equations
(\ref{eqn:code_cgs}) and (\ref{eqn:code_cgs_L}), we can investigate
the coronal properties of different accretion states. For a given
point in the corona, 
$\mathcal{L}$, $n_e$, and $U_{\rm ph}$ all scale linearly with
$\dot{m}$, so from equation (\ref{eqn:P_compt}) one can see that the
term $\gamma^2 \beta^2$ should scale like $\dot{m}^{-1}$. At low
electron temperatures we have
\begin{equation}
\gamma^2 \beta^2 \approx \frac{v^2}{c^2} \approx 3\frac{k_B T_e}{m_e
  c^2}\, ,
\end{equation}
while in the relativistic regime,
\begin{equation}
\gamma^2 \beta^2 \approx \gamma^2 \approx 12 \left(\frac{k_B
  T_e}{m_e c^2}\right)^2\, ,
\end{equation}
recovering the well-known scaling of IC power with temperature
\citep{rybicki:04}. Thus, at a fixed height above the disk, the
electron temperature should scale like $T_e \sim
\dot{m}^{-1}$ for $k_B T_e \ll m_e c^2$, while at high temperatures, $T_e \sim
\dot{m}^{-1/2}$, {\it independent of the black hole mass}.

However, since we have fixed the total coronal optical depth at unity,
the characteristic density near the photosphere is $n_e \simeq
(\sigma_T H_{\rm dens})^{-1}$, {\it regardless of $\dot{m}$}.  Therefore, at a
fixed optical depth $\tau$ in the corona, equation (\ref{eqn:P_compt}) becomes
\begin{equation}\label{eqn:T_LU}
\gamma^2 \beta^2 \simeq
\frac{3}{4}\frac{H_{\rm dens}}{c}\frac{\mathcal{L}}{U_{\rm ph}}\tau^{-1} \, .
\end{equation}
A rough model for how the temperature scales with $\dot{m}$
can then be derived if we treat both the density and the
dissipation profiles as exponentials with vertical scale heights
$H_{\rm dens}$ and $H_{\rm diss}$, respectively.  Because both
scale linearly with $\dot{m}$ [see eqns.~(\ref{eqn:code_cgs}) and
(\ref{eqn:code_cgs_L})], these two quantities can be described by
$\mathcal{L}(z) = \dot{m} \mathcal{L}_0 e^{-z/H_{\rm diss}}$ and
$\rho(z) = \dot{m} \rho_0 e^{-z/H_{\rm dens}}$. From the simulation data we find
that $H_{\rm diss} \simeq 3 H_{\rm dens}$.  It then follows that, at
fixed optical depth $\tau$, 
$\mathcal{L}$ will scale with $\dot{m}$ like
$\mathcal{L} \sim
\dot{m}^{(1-H_{\rm dens}/H_{\rm diss})} \sim \dot{m}^{2/3}$ (this also
explains the scaling of $L_{\rm cor}/L_{\rm tot} \sim \dot{m}^{-1/3}$
seen in Table \ref{table:L_cor}). 
On the other hand, $U_{\rm ph}$ is nearly constant throughout the corona,
but scales linearly with $\dot{m}$, so equation (\ref{eqn:T_LU}) becomes
\begin{equation}\label{eqn:Tscale}
\gamma^2 \beta^2 \sim
\frac{\mathcal{L}}{U_{\rm ph}}\tau^{-1} \sim \dot{m}^{-1/3} \tau^{-1}\, .
\end{equation}
We therefore expect the temperature {\it at a fixed optical depth} to
scale like $T_e \sim \dot{m}^{-1/3}$ in the non-relativistic regime,
and $T_e \sim \dot{m}^{-1/6}$ at high temperature.
Similarly, at fixed $\dot{m}$, we expect $T_e \propto \tau^{-1}$
non-relativistically and $T_e \propto \tau^{-1/2}$ when the electron
temperature is relativistic.

\begin{figure}
\caption{\label{fig:Te_r} Mean coronal temperature as a function of
  radius, weighted by local cooling rate $\mathcal{L}$ (top) and
  electron density $n_e$ (bottom), for a range of luminosities.} 
\begin{center}
\includegraphics[width=0.9\linewidth]{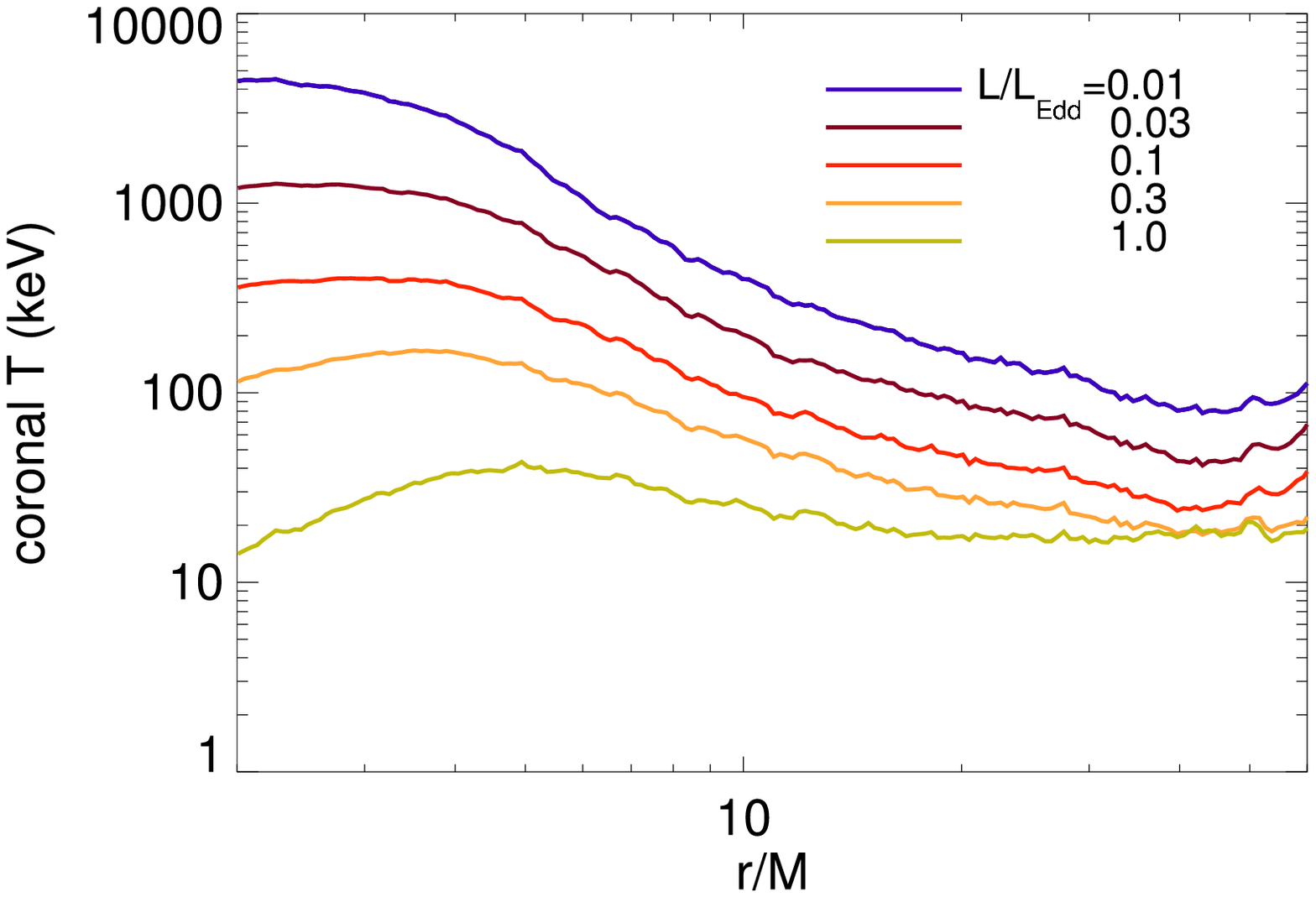}\\
\includegraphics[width=0.9\linewidth]{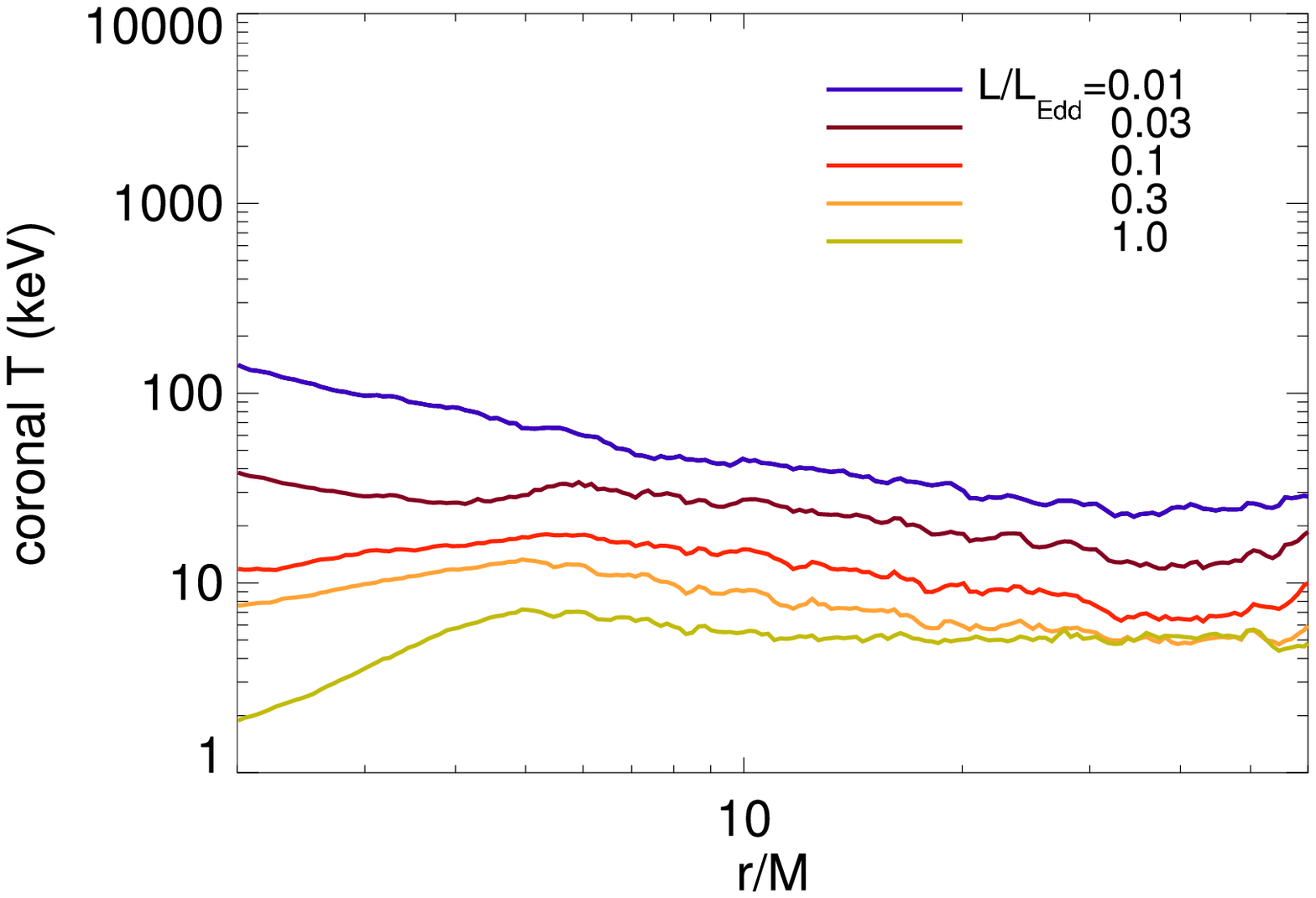}
\end{center}
\end{figure}

Thus, whether comparing regions of constant latitude or constant
optical depth, we
see a clear trend that is consistent with decades of
observations: low-luminosity states are characterized by hard X-ray
flux from a hot corona, while high-luminosity states lead to a much
cooler corona and softer spectrum. In Figure \ref{fig:Te_r} we plot the
time-averaged coronal
temperature as a function of radius for a range of different accretion
rates. In the top panel the mean temperature is calculated by
integrating over $\theta$ and $\phi$ and weighting by the local cooling
rate $\mathcal{L}$, while in the bottom panel the temperature is
weighted by the electron density $n_e$. The $\mathcal{L}$-weighting
is more closely related to the emergent spectrum and
naturally probes the upper corona, while the $n_e$-weighting speaks
to conditions in the majority of the coronal mass and is
sensitive to the conditions near the disk. In either case, the trend
with $\dot{m}$ is clear and, at the level of approximation expected,
consistent with our earlier rough scaling argument. We also see that
in the bulk of the corona, especially outside the ISCO, the temperature
changes very little with radius.

\begin{figure}[ht]
\caption{\label{fig:Te_tau} Mean coronal temperature as a function of
  optical depth at a range of radii, for $L=0.1L_{\rm Edd}$.}
\begin{center}
\includegraphics[width=0.9\linewidth]{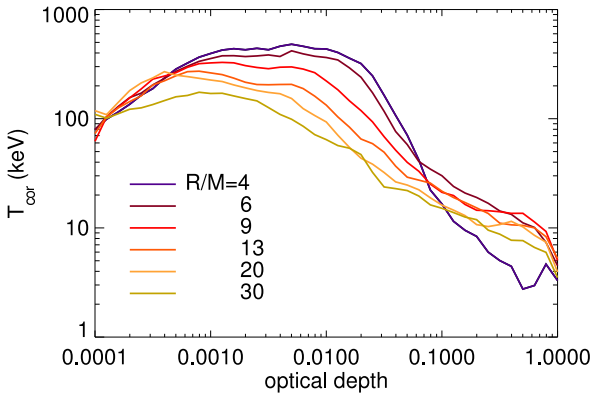}
\end{center}
\end{figure}

The time-averaged radial and vertical temperature profiles of the corona
can be seen in greater detail in Figure \ref{fig:Te_tau} for $\dot{m}=0.1$.
At six different values of $r$, we plot the temperature as a function of optical
depth through the corona, where $\tau=0$ corresponds to the $z$-axis, and $\tau=1$
the disk surface (see contours of $\tau$ in Fig.\ \ref{fig:Te_contour}).
From the base of the corona at $\tau=1$ outward to $\tau \sim 0.01$,
the predicted $\tau^{-1}$ scaling describes the results well for all
radii outside the ISCO.
Between $\tau=1$ and $\tau=0.1$, where most of the
scattering events occur, and roughly $50\%$ of the coronal cooling
takes place, the temperature is always between 3 and 20~keV.
This is a relatively low temperature for a disk corona, resembling 
more a warm atmosphere than a hot corona. Only in the upper corona,
where the density and optical depth are least, does the temperature
surpass 100 keV. Yet, because it accounts for the other $50\%$ of
the coronal cooling, even
this small amount of hot gas is sufficient to contribute a hard 
power-law tail to the X-ray continuum.

Another interesting feature seen in Figure \ref{fig:Te_tau} is the
turn-over in temperature for $\tau \lesssim 10^{-3}$. As can be seen
from the temperature map in Figure \ref{fig:Te_contour}, this region
is very close to the funnel/jet region, where significant outflows are
expected. Since \harm cools only {\it bound} matter, $\mathcal{L}$ is
set to zero for much of this region, leading to a decreased average
temperature.  This is, of course, an artifact of the simulation. Because
the black hole in this simulation does not rotate, the jet power is
very small and this artifact should be unimportant; when the
black hole rotates and the jet power is greater, the dissipation rate
in the jet could be significant.  At the same time, the large relativistic bulk motion of
gas in this region can still lead to interesting Comptonization
effects, as will be described below in Section \ref{section:bulk}. 

Finally, we comment on the relationship between our calculated
electron temperature $T_e$ and the nominal gas temperature $T$ found in the simulation.
In a fully self-consistent picture, we would expect $T_e$ to be close
to the ion temperature $T_i$ because the coronal densities are generally high
enough to make ion-electron collisional coupling reasonably rapid: just above
the photosphere, the electron heating rate due to Coulomb collisions with
hotter ions is
\begin{equation}\label{eqn:coulomb}
\frac{d}{dt}\ln (3/2 kT_e) \simeq 0.05
           \frac{g(T_e)}{4}\left(\frac{H_{\rm dens}/r}{0.06}\right)^{-1}
           \left(\frac{r}{10M}\right)^{-1}\left(T_i/T_e - 1\right) M^{-1},
\end{equation}
where we have estimated the local electron density by $\simeq
(\sigma_T H_{\rm dens})^{-1}$.
When the ion temperature is $\ll m_p c^2$, the function $g(T_e)$ falls sharply
with increasing temperature below $\simeq 50$~keV, but varies slowly when $T_e \gtrsim 100$~keV.
Consequently, for most of the range of interest, $g(T_e) \simeq 4$ \citep{stepney:83}.
The estimate of equation (\ref{eqn:coulomb}) then shows that the
thermal equilibration time is at most comparable to, and in much of the corona
considerably shorter than, the dynamical time $\simeq 32
(r/10M)^{3/2}M$, so treating it as a single-temperature fluid should
be a good approximation.

In practice, we find that the ratio $T_e/T$ is generally considerably less than
unity.   For example, if $\dot m = 0.1$, it ranges from $\sim 5 \times 10^{-3}$
(very near the photosphere) to $\sim 0.2$ (in the hottest locations at high altitude).
For other values of $\dot m$, this ratio will scale like $T_e$ because $T$ is a fixed
property of the simulation.  The actual gas pressure in the corona is therefore
likely overestimated in the simulation.  Fortunately, however, it has little
influence on the density structure of the corona.  It is smaller than both the
radiation pressure and the magnetic pressure by factors of $\sim 3$--30.  Replacing
the code temperature with $T_e$ would therefore make the gas pressure even less
significant in coronal dynamics.

\section{BROAD-BAND SPECTRA}\label{section:broad_band}

Having converged on a self-consistent, global map of the coronal
electron temperature, there is little left to do but ``turn the
crank'' with \pand, ray-tracing as many photon packets as
computationally reasonable. As described in \citet{schnittman:12}, the
photon packets are emitted from the photosphere of the disk with a
(diluted) thermal spectrum, and subsequently up-scattered via inverse Compton in the
corona, eventually either getting captured by the black hole or reaching an
observer at infinity. Those photons that escape are binned by their
energy and observer coordinates $(\theta, \phi)$, making it trivial to
generate simulated X-ray spectra as a function of viewing angle. Since
the \harm data is fundamentally dynamic, it is also straight-forward
to simulate X-ray light curves and investigate timing features such as
quasi-periodic oscillations and time lags between hard and soft
bands. An in-depth study of these topics will be the subject of a
future paper, but in this work, we generally average the spectra over
multiple snapshots.

Specifically, we use ThinHR simulation data from snapshots between
$10000M$ and $15000M$, sampled every $500M$ in time. Our photon packets cover
the range of energy from $10^{-3}$ to $10^4$ keV, with logarithmic
spacing and spectral resolution of $\Delta E/E=0.016$. By using multi-energy
photon packets, we are able to resolve the thermal continuum with high
accuracy and efficiency. However, the Monte Carlo IC scattering kernel
still introduces a significant amount of numerical noise at high
energies \citep{schnittman:12}. This noise can be reduced by
increasing the number of rays traced, but in practice seems to
converge only slowly.

\begin{figure}[ht]
\caption{\label{fig:spectra_lum} Broad-band X-ray spectra for
  $M=10M_\odot$ and a range of
  luminosities, integrated over all viewer inclination angles. In each case,
  the spectrum includes a broad thermal peak around 1-3 keV, a
  power-law tail and Compton reflection hump above 10 keV, and a broad
  iron line at 5-7 keV. The sharp lines above 30 keV are due solely to
  Monte Carlo fluctuations.} 
\begin{center}
\includegraphics[width=0.9\linewidth]{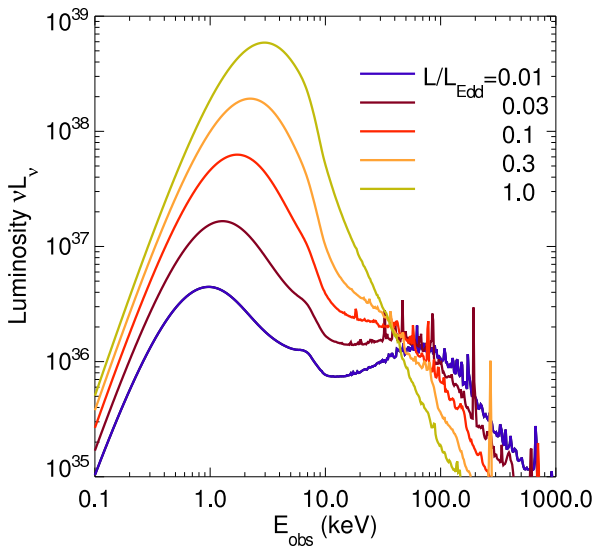}
\end{center}
\end{figure}

Figure \ref{fig:spectra_lum} shows the observed spectra from ThinHR
for a range of $\dot{m}$, integrated over all viewing angles. The
dominant features include a broad thermal peak around 1--3 keV, a
power-law tail, and a Compton reflection hump above 10 keV.\footnote{Because
we do not yet include photoionization losses other than Fe K-shell
ionization, reflection at energies below 7~keV is not suppressed.
At the temperatures characteristic of the inner regions of accretion
disks around stellar-mass black holes, this should be a reasonable
approximation, although it fails for AGN.}
There is no evidence for a cutoff up to at least 1000 keV, but the
Monte Carlo statistics are very poor at those high
energies
%%JDS
(we do not currently include reprocessing of energy lost by photons
above $\sim 100$ keV due to Compton recoil in the disk, although from
Fig.\ \ref{fig:spectra_lum} this is clearly a small fraction of the
total energy budget).
Also visible is a broadened iron line feature around 5--7
keV, which will be discussed in greater detail in the following
section. 

The most important result to be seen in Figure \ref{fig:spectra_lum}
is that, {\it for the first time, we have been able to use the genuine
physics of global MHD simulations
to reproduce the X-ray spectra observed in a wide variety of
black hole binary states.}
%Moreover, we are able to do so even while
%retaining an optically thick thermal disk extending to small radii.}
%In Table \ref{table:states} we give a summary of the spectra plotted
%in Figure \ref{fig:spectra_lum}, 
%%%JDS
We carry out a simple phenomenological fit of the spectra in Figure
\ref{fig:spectra_lum} using a fully relativistic multi-color disk, a
gaussian peak around 6 keV, and a power law component at higher
energies. The resulting best-fit parameters are summarized in Table
\ref{table:states}, using the classifications defined by
\citet{remillard:06}. The disk fraction $f^b$ is limited to the 
2--20 keV band, as in \citet{remillard:06}.
Note that $f^b$ is (particularly for small
$\dot m$) smaller than $1 - L_{\rm cor}/L_{\rm tot}$ shown in Table~\ref{table:L_cor}
because it is the fraction only within the 2--20~keV band, and much
of the disk power is radiated at lower energies. The power-law
index $\Gamma$ is taken from the number flux of photons per unit
energy $N(E) \propto E^{-\Gamma}$. While we do not claim to completely
fit the spectra with only a thermal peak, gaussian line, and a single
power-law tail, $f^b$ and $\Gamma$ are still valuable parameters for
spectral classification. 
%%%

\begin{table}[ht]
\caption{\label{table:states} Broad-band spectral properties for a
  range of mass accretion rates $\dot{m}$. The disk has a peak
  temperature $T_{\rm disk}$ and contributes a fraction $f^b$ to the
  total flux in the 2--20 keV band. The power-law index $\Gamma$ is
  measured between 10 and 100 keV, and the state corresponds to the
  classification of \citet{remillard:06}.}
\begin{center}
\begin{tabular}{lcccc}
\hline
\hline
$\dot{m}$ & $k T_{\rm disk}$ (keV) & $f^b$ & $\Gamma$ & state \\
\hline
0.01 & 0.42 & 0.19 & 1.6 & hard \\
0.03 & 0.54 & 0.41 & 2.0 & hard/SPL \\
0.1  & 0.64 & 0.67 & 2.6 & SPL \\
0.3  & 0.81 & 0.82 & 3.1 & SPL/thermal \\
1.0  & 1.09 & 0.90 & 4.0 & thermal \\
\end{tabular}
\end{center}
\end{table}

When scaling the simulations to $\dot{m}=0.01$, we reproduce 
%%JDS
some of
the features that characterize the 
low-hard state described in \citet{remillard:06}, with
$\Gamma < 2.1$ and the 2--20 keV flux dominated by the corona: $f^b
< 0.2$. At $\dot{m}=0.1$ and above, the spectra closely resemble observations of
the steep power-law (SPL) state, with $\Gamma > 2.4$ and a disk
contribution of $0.2 < f^b <0.8$. At the highest luminosities, we are
most closely aligned with
the thermal state, defined by $f^b >0.75$ and little
variability (see below, Sec.\ \ref{section:variability}). 

Despite the remarkable success of reproducing such a wide range of
spectral behavior with a single simulation, we should note that these
spectra represent just a one-dimensional slice through the
hardness-luminosity plane that is populated by stellar-mass black
holes with a wide diversity of behaviors. For example, LMC X-3 alone
has been observed with $\dot{m}$ anywhere from $<0.03$ up to $>0.5$
{\it in the thermal state alone} \citep{steiner:10}. Furthermore, with
our current techniques, we are not able to reproduce the very
pure thermal spectra used for inferring spin with the continuum
fitting technique \citep{mcclintock:06}. 
A complete description of the spectral states of black hole binaries
may also require additional parameters such as black hole spin and
magnetic field topology.  For example, a magnetic field that is
primarily toroidal may curb coronal activity \citep{beckwith:08}.
Exploration of how accretion rate and these additional parameters
interact will be the topic of future simulations.

%%JDS substantial re-write
We also note that the 
thermal disk component in the $\dot{m}=0.03$ case is somewhat larger than that
traditionally inferred in the low-hard state. 
In part, this soft component is due directly to the thermal disk, but
much of it is also due to the disk photons that get upscattered in the
warm, high density regions ($T_e \approx 10$ keV; $\tau
\gtrsim 0.1$) of the corona. This effect is evident in the Wien tails
of the thermal peaks in Figure \ref{fig:spectra_lum}, which are
noticably harder for smaller values of $\dot{m}$ due to the higher
coronal temperatures in those cases.
In fact, for some observations of the hard state,
more recent analyses
have shown clear evidence for an optically thick thermal disk
\citep{miller:06a,miller:06b,reis:10,hiemstra:11}. In those
observations, the disk extends in to the ISCO with a relatively cool
temperature of $\sim 0.2$--0.35 keV, 
slightly lower than we find for $\dot{m}\le 0.03$. 
%Clearly the canonical
%disk+corona geometry for black hole accretion can extend even to the
%low-hard state in many cases, where a substantial fraction of thermal
%disk photons can get efficiently up-scattered to create a hard
%spectrum. 
On the other hand, there are also observations of the hard state that
show evidence of a disk truncated at large radius \citep{esin:01,done:07}.
It is possible that the inner regions of these systems may be dominated
by a radiatively inefficient flow (e.g., \citet{esin:97}), and thus are
not well-represented by this simulation.
As with the thermal state, we are not able to reproduce
every observed black hole spectrum simply by varying $\dot{m}$ within
a single simulation.  

The key elements in our model that lead to the shift in spectral shape
with accretion rate are the assumptions that the gas density scales
linearly with $\dot{m}$ (discussed below in Sec.\
\ref{section:disk_theory}) and that the boundary of the
corona is defined by the disk's scattering photosphere.  The latter
assumption is physically reasonable: as shown in our calculation,
the coronal temperature declines close to the disk surface; in
addition, the quickly rising density there leads to a much greater
importance for cooling processes like bremsstrahlung. 
For fixed $H_{\rm dens}$, the natural consequence of these assumptions
is that as $\dot m$ increases, a larger fraction of 
the dissipation takes place inside the thermal disk and a smaller
fraction in the corona.  The disk appears weaker at low accretion rates
because the corona, although only marginally thick by assumption,
can give so much energy to the average photon created thermally
in the disk.  In Section~\ref{section:disk_theory},
we will discuss how a more realistic disk picture, in which
$H_{\rm dens}$ changes with $\dot m$, may affect our spectral predictions.

We have also investigated the effect of observer orientation on the
shape of the spectrum, but in most cases find only weak dependence on
inclination. After accounting for projection effects, we do see that 
high-inclination (i.e., edge-on) systems have a smaller thermal peak,
and somewhat harder spectrum between 1 and 10 keV, consistent with the
results found in \citet{noble:11}. Above $\sim 30$ keV, the spectra
are virtually identical. This is quite reasonable considering the
results of the previous section, where we showed the coronal
temperature increasing significantly with distance above the disk. The
highest-energy photons are mostly generated in a large, diffuse volume
in the upper corona, which subtends roughly the same solid angle
independent of viewer inclination.
The only part of the spectrum that seems to be strongly sensitive to
the inclination is the broad iron line, which will be
described in the following section.

\begin{figure}[ht]
\caption{\label{fig:spectra_r} Broad-band spectra decomposed into
  relative contributions from different radii in the disk. From top to
  bottom, $L/L_{\rm Edd}=0.01, 0.1, 1.0$.}
\begin{center}
\includegraphics[width=0.9\linewidth]{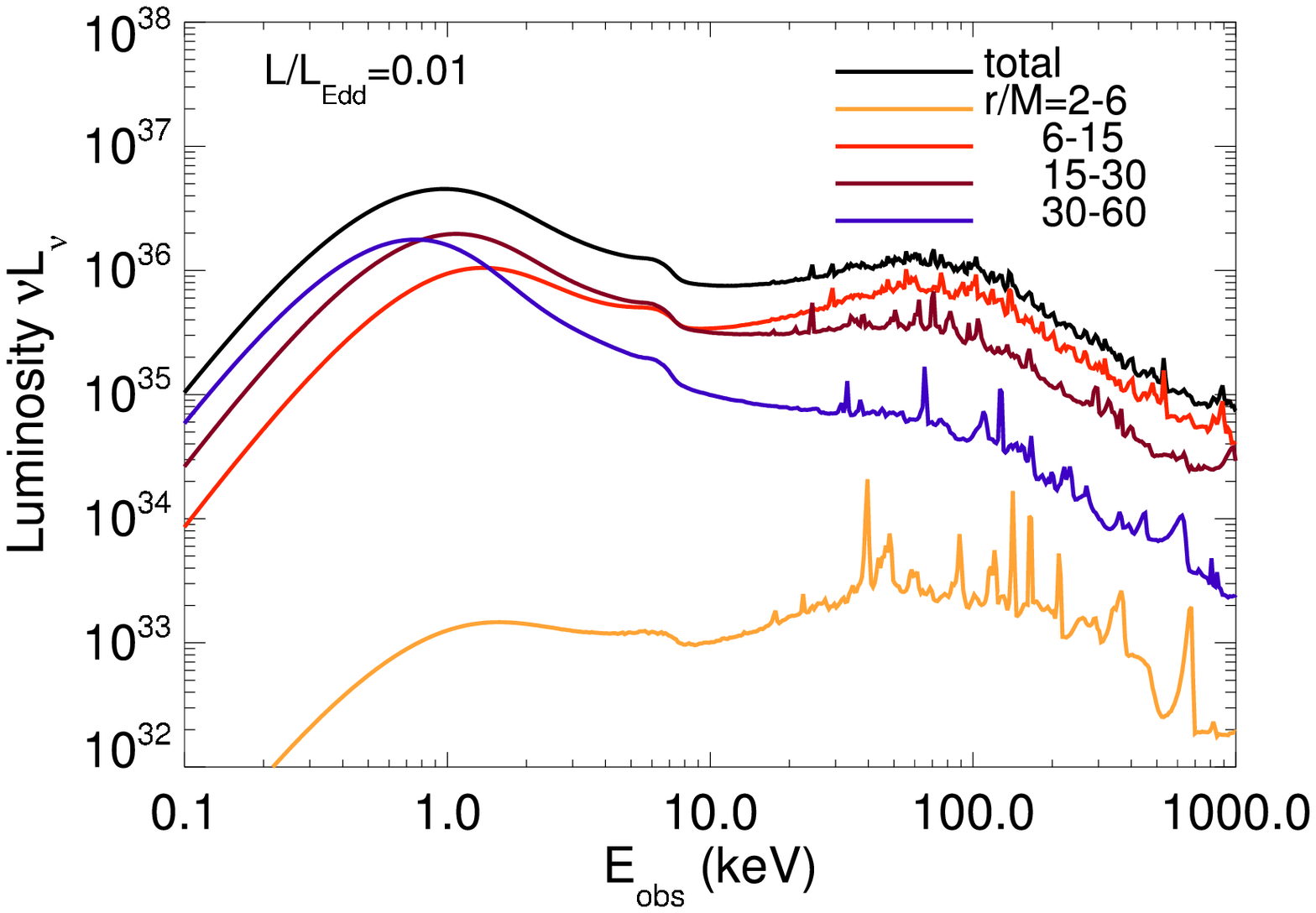}
\includegraphics[width=0.9\linewidth]{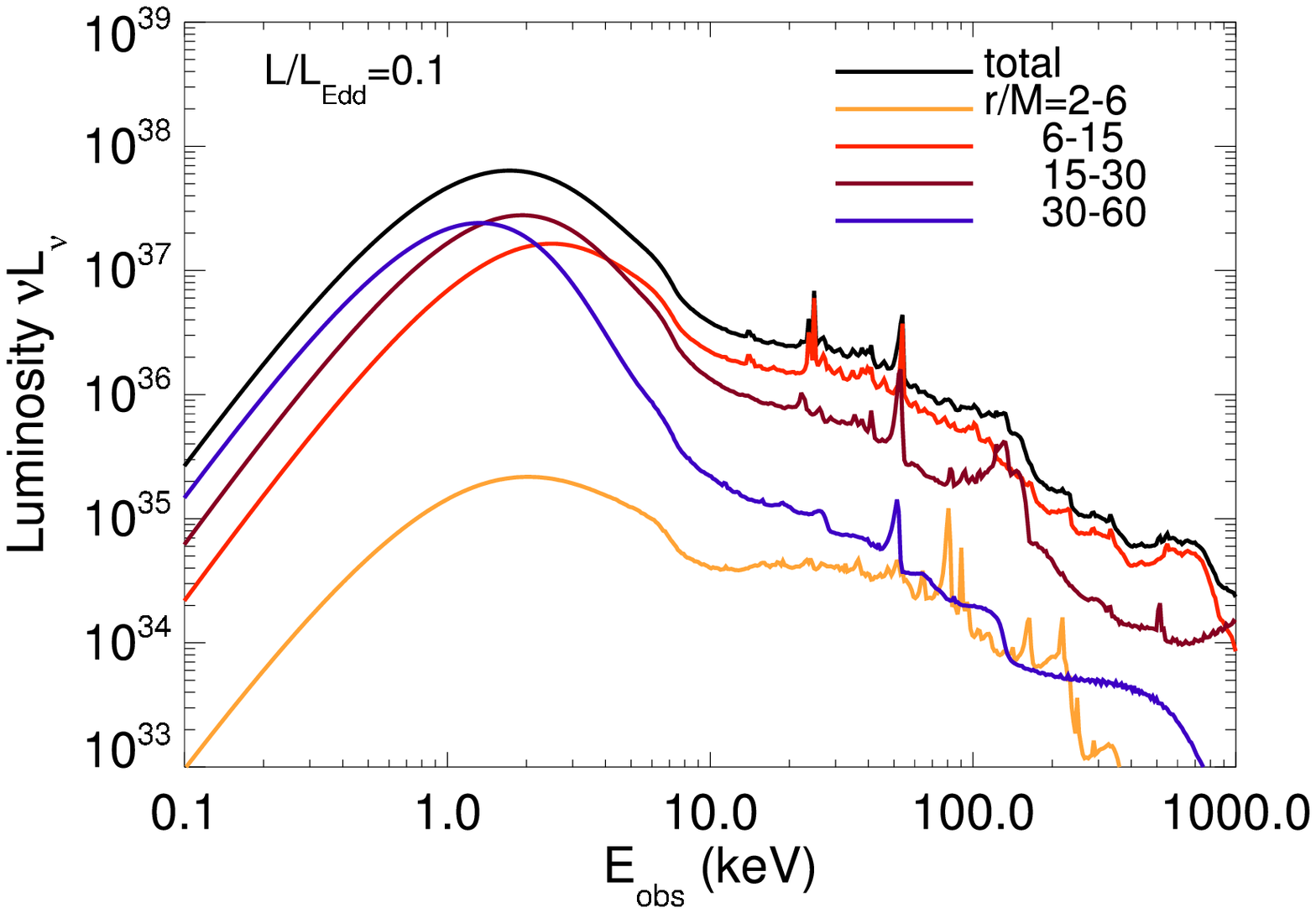}
\includegraphics[width=0.9\linewidth]{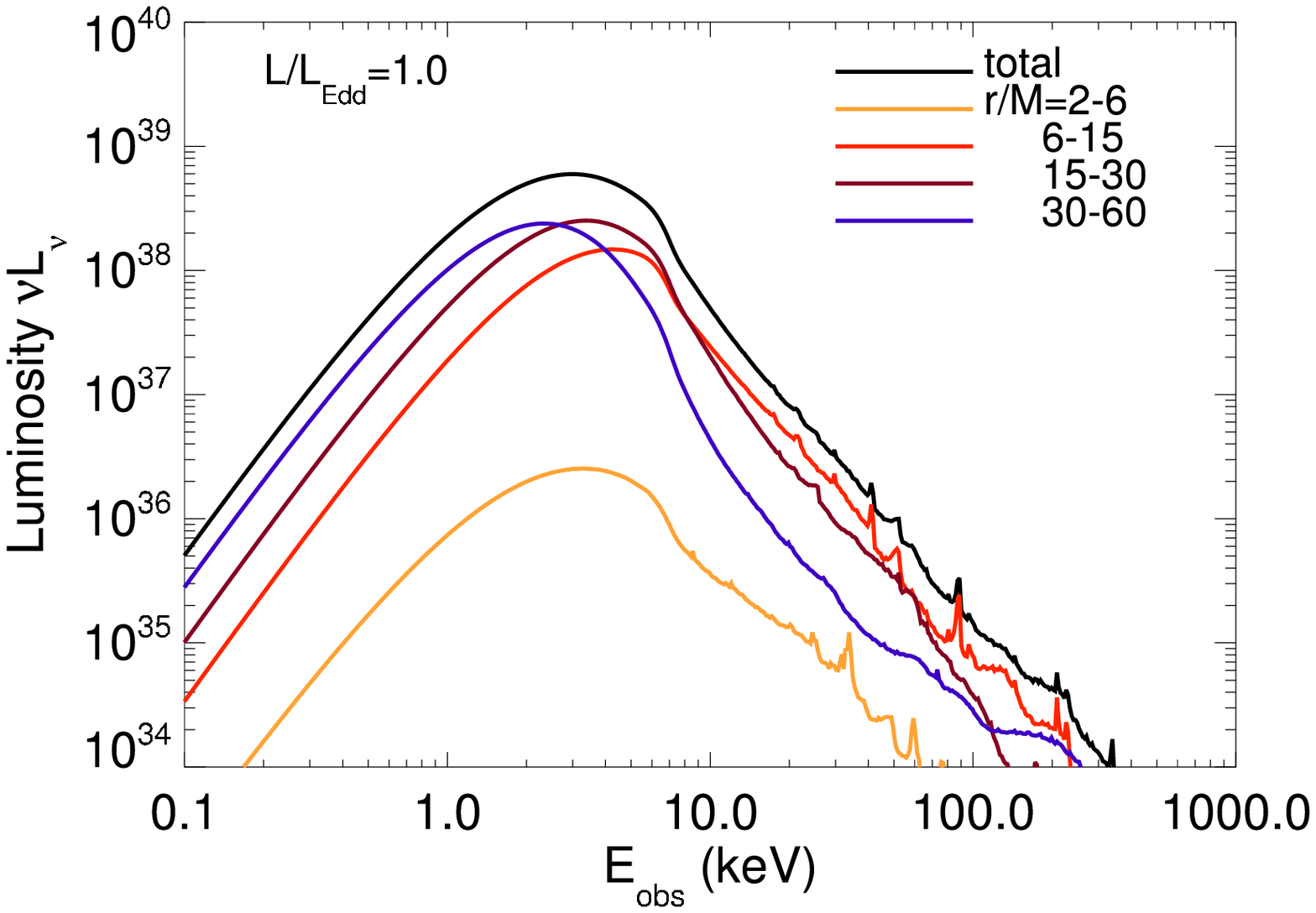}
\end{center}
\end{figure}

In Figure \ref{fig:spectra_r} we plot the broad-band spectra for
$\dot{m}=0.01$, 0.1, and 1.0, showing the relative contributions 
from different regions of the accretion disk. The spectra are sorted
by the emission radius of the seed photon. We see a few clear trends
across all accretion rates, none of which is very surprising: the
spectra grow systematically softer
with increasing radius, the thermal emission is dominated by flux
originating from $r/M>15$, the coronal emission is dominated by the
region $6<r/M<15$, and the plunging region contributes very little to
either the thermal or power law parts of the spectrum. 

The fact that the plunging region contributes so little to the spectrum
does not mean that the classical N-T disk is an adequate model for
accretion dynamics. As shown in \citet{noble:11} and
\citet{kulkarni:11}, the radial emissivity profile from MHD
simulations leads to thermal spectra that are systematically harder
than Novikov-Thorne would predict for the same spin parameter. In
part, this comes from the small amount of
dissipation from inside the ISCO, but an even more important cause is
the emission profile immediately outside the ISCO, which peaks at a
smaller radius than predicted by Novikov-Thorne, and thus the
MHD thermal spectra look like they come from black holes with somewhat higher
spins \citep{noble:11,kulkarni:11}.

\begin{figure}[ht]
\caption{\label{fig:spectra_tau} Broad-band X-ray spectra for
  $\dot{m}=0.1$, when varying the optical depth of the
  photosphere. While the total fraction of hard X-ray flux is
  directly proportional to the total fraction of dissipation in the corona, the
  shape of the spectrum appears to be largely independent of this
  parameter.}
\begin{center}
\includegraphics[width=0.9\linewidth]{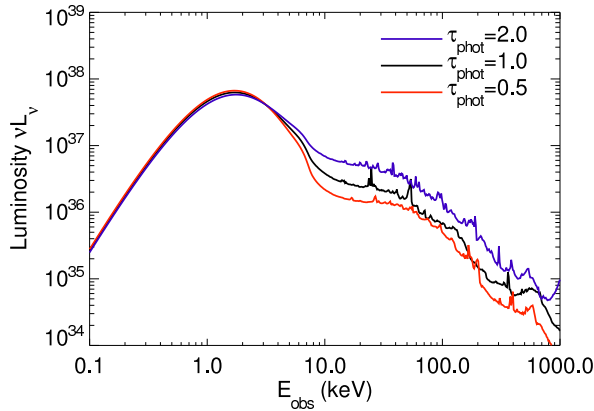}
\end{center}
\end{figure}

One of the major underlying assumptions of this paper is that
the disk photosphere is placed at the $\tau_{\rm phot}=1$ surface. We believe
this is an eminently reasonable and physically-motivated assumption,
but it is worth investigating how much our central results are sensitive
to it.  To this end, we have repeated
the entire radiation-temperature iterative solution for $\dot{m}=0.1$, setting
$\tau_{\rm phot}=0.5$ and again with $\tau_{\rm phot}=2.0$. The results are shown
in Figure \ref{fig:spectra_tau}. While the total luminosity is of
course unchanged, the relative flux in the hard X-ray tail
necessarily increases with $\tau_{\rm phot}$. However, while the
normalization of this hard tail changes, the {\it shape} appears to be
invariant, indicative of an identical temperature profile in the upper
corona, regardless of the exact location of the photosphere. This
follows from the basic nature of inverse Compton radiation: equation
(\ref{eqn:P_compt}) shows that the electron temperature is set by the
total radiation density, not the spectrum 
(at very high photon
energies, pair creation and relativistic corrections to the Thomson
scattering cross section will be required; we have not included them
in this treatment, which is reasonable considering the
rather small fraction of the total luminosity with energy $> 100$~keV).
So the energy balance in 
the upper corona is completely insensitive to the detailed radiative
processes taking place in the disk and boundary layer. 

\section{IRON EMISSION LINES}\label{section:iron_lines}

Relativistically broadened Fe K$\alpha$ lines have been detected
in numerous AGN \citep{tanaka:95,nandra:07,brenneman:09}, galactic black holes
\citep{miller:04,reis:08, reis:09}, and galactic neutron stars \citep{cackett:10}.
The underlying emissivity profile is nearly always inferred
(e.g., \citet{reynolds:03}) by fitting the observed line profile to a phenomenological
model in which the emissivity is zero inside the ISCO, rises
abruptly to a maximum at the ISCO, and then declines as a power-law
(sometimes a broken power-law) toward larger radii.  The energy with which
fluorescence photons arrive at a distant observer depends on
the radius from which they are emitted and the direction in
which they are sent, as well as the character of the spacetime
in which they travel.
It would, of course, be highly desirable both to find functional
forms for the K$\alpha$ emissivity that are more closely connected to
physical considerations and to be able to use observational data to
constrain the disk dynamics responsible for generating these lines.

To do so requires solving problems both of physics and of procedure.
Fluorescence line production begins with the illumination of
gas by X-rays of energy greater than the threshold for K-shell
ionization; we must determine its intensity as a function of
radius.   The fraction of those photons absorbed by such
ionization events depends on the total optical depth of the
gas and the ratio between the absorption opacity and other opacities
(predominantly Compton scattering).  The total optical depth
depends on the specifics of angular momentum transport within
the disk.  The absorption opacity
(as well as the fluorescence yield and the line energy) depend
on the ionization state of the Fe atoms, and that in turn
depends on both the temperature in the absorbing layer and
the ratio between the ionization rate and the recombination
rate.  Although relativistic ray-tracing in vacuum has long
been a solved problem \citep{carter:68,bardeen:72}, a significant fraction of
K$\alpha$ photons traversing a marginally optically thick corona
may also gain or lose energy by Compton scattering.  A significant
procedural problem is posed by the question of how to separate line
photons from the continuum \citep{miller:07}. This is particularly
problematic in the case of stellar-mass black holes, where the thermal
peak and the power-law tail intersect right around the iron line,
making it challenging to determine the precise form of the underlying continuum spectrum.

Our new ray-tracing analysis of the \harm simulations directly solves
many of these problems.  The radial profile of hard X-ray illumination
is a direct product of our global solution for the radiation field.  The total optical
depth of the disk is automatically computed by the underlying
general relativistic MHD simulation, subject only to scaling with
our choice of $\dot m$.  Compton scattering {\it en
route} also follows naturally from our Monte Carlo transfer
solution. Even the continuum contribution is also an automatic
byproduct, greatly improving our ability to uniquely fit the shape of
the iron line.

The principal remaining uncertainty is calculation of the ionization
state. In this paper we assume a fixed ionization state, but the data
required for a genuine calculation of the ionization state as a
function of position are also supplied by the other components of our
method, so even this last problem can be solved within our framework,
although it will involve a certain amount of additional labor.

As the disk seed photons are scattered through the corona, many eventually
return, with higher energy, to the disk photosphere. For stellar-mass black
holes with disk temperatures of $\sim 1$ keV, essentially all Fe
atoms will be ionized to only a few remaining electrons.  However,
the ability to produce a K$\alpha$ photon as a result of K-shell
photoionization disappears only when the Fe is completely stripped.
From Saha equilibrium, we find that most of the photosphere is dominated
by He-like Fe~XXV for $\dot{m} \le 0.1$, but a mix of He-like, H-like, and
fully stripped Fe exists for $\dot{m} \ge 0.3$.

\begin{figure}[ht]
\caption{\label{fig:ionizing_flux} Absorbed iron K-edge photon flux in
  the local fluid frame, assuming a uniform ionization state. For lower
  luminosities, the disk becomes optically thin outside the horizon, leading
  to a clear turnover and cutoff of K-shell excitation in the plunging
  region. Also shown (dashed line) is the outgoing seed photon flux
  $F_{\rm em}(E>3\mbox{ keV})$ 
  for $\dot{m}=0.1$, normalized to appear on the same axes.}
\begin{center}
\includegraphics[width=0.9\linewidth]{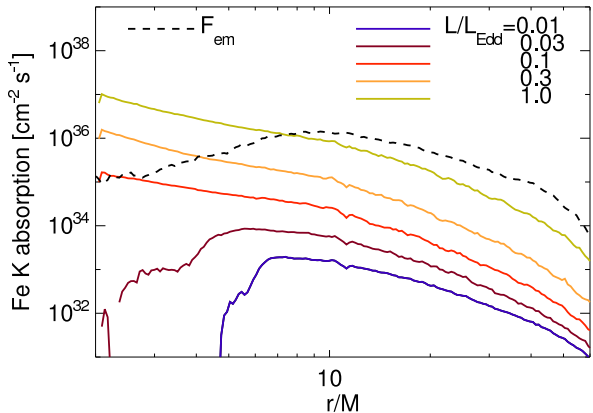}
\end{center}
\end{figure}

The K-edge threshold varies slowly with ionization state, from $\sim
7$ keV for neutral iron, up to 8.8 keV for Fe~XXV and 9.3 keV for Fe~XXVI
\citep{kallman:04}.  At photon
energies much above this threshold, the photoionization cross section decreases
sharply with energy. In Figure \ref{fig:ionizing_flux} we show the
radial profile of the absorbed K-edge photon flux, defined as the 
incident photon number flux in the 9--30 keV band times the fraction
of the disk that is optically thick at that radius\footnote{The
  reflection edge of the disk is not a sharp boundary; at a given
  radius, the optical depth as a function of azimuth and time can
  change by a factor of a few.}. For
comparison, for $\dot{m}=0.1$, we also show (dashed line) the number
flux of seed photons
emitted from the disk $F_{\rm em}$ with energy greater than $\sim 3$ keV, i.e.,
those most likely to get up-scattered to $>9$ keV. Note that this plot
of the outgoing flux is normalized for comparison purposes. 

In the outer regions of the disk, the absorbed K-edge flux profile is similar
in shape to the emitted flux of photons above 3~keV.
However, in the plunging region the disk becomes optically thin when
$\dot m < 0.1$ (see Table~\ref{table:L_cor}), suppressing the
absorption there.
For $\dot{m}\gtrsim 0.1$, the optically thick disk extends all the
way to the horizon, and the shape of the absorbed flux profile is
essentially independent of accretion rate. 
Even as the disk emission falls off with
smaller radius, the illumination profile continues to rise as $F_{\rm Fe}
\sim r^{-\alpha}$, with $\alpha\approx 3/2$,
indicative of the increasing importance of coronal
flux in the inner disk. 

We see no evidence for a {\it steepening} of the radial illumination
profile with decreasing radius, as suggested by some AGN observations
\citep{vaughan:04,miniutti:07,wilkins:11,ponti:10} and
predicted by ``lamp post'' models \citep{george:91}, in which
the hard flux comes from a concentrated region along the black hole
axis. This is not very surprising, considering the density and
luminosity maps in Figures \ref{fig:rho_contour} and \ref{fig:L_contour},
which show an evacuated funnel around $\theta = 0$, essentially the
opposite of the lamp post geometry.  Indeed, a stationary point source
on the rotation axis seems rather unlikely dynamically: a centrifugal
barrier prevents much matter from approaching close to the axis,
and any matter with little enough angular momentum to enter that
region must either fall rapidly into the black hole or be ejected;
in both cases, there would be strong beaming of any photons emitted
in the direction of travel. It is possible that non-zero black hole spin
may lead to relatively greater coronal dissipation at small radii, removing this
discrepancy with observations; future simulations can test this conjecture.

As mentioned above, the line profile is also sensitive to the iron
ionization state as a function of radius. Even if the surface density
of the disk remains large inside of the ISCO, a line can be produced only
if the iron is not fully ionized \citep{reynolds:97}. 
\citet{reynolds:08} initiated the use of simulation data to
predict Fe Kalpha profiles by placing a source of ionizing radiation
on the rotation axis $6M$ above the disk, and then using density
data from a pseudo-Newtonian MHD simulation to estimate the ionization
parameter in the disk.  Since we know the vertical
density profile as well as the incident spectrum at each point in
the disk, it should be possible to completely solve the ionization
balance equations as in \citet{garcia:11a} and \citet{garcia:11b}. Such a
detailed treatment is beyond the scope of the present work, but we can
substitute reasonable approximations to obtain useful first-order
results.

When a photon packet hits the disk photosphere, some part is
absorbed by the iron atoms, while the remainder is reflected by electron
scattering (other processes, such as free-free absorption in the disk,
are insignificant). For a single photon incident on the disk,
the probability of absorbing the photon in Fe K-shell photoionization is
\begin{equation}
P(E) = \frac{N_{\rm scat} \kappa_{K\alpha}(E)}
{N_{\rm scat}\kappa_{K\alpha}(E)+\kappa_T}\, , 
\end{equation}
where $\kappa_{K\alpha}$ and $\kappa_T$ are the Fe K$\alpha$ and
Thomson scattering opacities. $N_{\rm scat}$ is the 
median number of scattering events a photon experiences before emerging from the
atmosphere. Thus, the typical photon gets $N_{\rm scat}$
chances to excite a K$\alpha$ transition before exiting the disk,
thereby enhancing the yield on the line production \citep{kallman:04}. For
accretion disks with roughly solar abundances and dominated by He-like
Fe, we take
\begin{eqnarray}
\kappa_{K\alpha}(E<8.8 \mbox{ keV}) &=& 0 \nonumber\\
\kappa_{K\alpha}(E>8.8 \mbox{ keV}) &=& \kappa_T \left(\frac{E}{8.8 \mbox{
    keV}}\right)^{-3} 
\end{eqnarray}
and from Monte Carlo scattering experiments, we find an angle-averaged
value of $N_{\rm scat}=3$.
This crude approximation to the K-shell opacity
is appropriate provided most Fe atoms retain at least one electron;
when most Fe atoms are stripped, $\kappa_{K\alpha}$ is smaller than
our estimate by the ratio of FeXXVI ions to the total.  Of all the
photons absorbed by iron in the disk, only a fraction $f_{K\alpha}$ produce
a fluorescent line, while the excitation energy deposited by the rest
is lost to Auger transitions, or, in the case of H-like and He-like Fe,
more energetic K series recombination lines \citep{kallman:04}.  In \pand's
current form, the energy absorbed by K-edge opacity is simply removed
from the spectrum during the Monte Carlo solution, while the energy in K$\alpha$
emission is added back later.  The fluorescence yield $f_{K\alpha}$ depends on ionization
state \citep{krolik:87}, growing slowly from $\simeq 0.34$ to $\simeq 0.5$
from FeI to FeXXII.  At higher ionization stages, it can be as little as 0.11
(FeXXIII), but is generally larger (0.5--0.75).  For all the results presented
below, we take $f_{K\alpha}=0.5$, corresponding to a highly ionized state. 

For a photon packet incident on the disk with initial spectral intensity $I_{\nu,0}$
[units of erg/s/Hz], the number of Fe K$\alpha$ photons produced per
second will be
\begin{equation}
N_{K\alpha} = f_{K\alpha} \int d\nu P(h\nu) \frac{I_{\nu,0}}{h\nu}\, .
\end{equation}
In our simplified model, all of these photons are added back to the
photon packet as a delta function in energy at $E=6.7$ keV,
corresponding to the K$\alpha$ emission for Fe~XXV.
Including absorption, the outgoing spectrum can be written
\begin{equation}
I_{\nu, \rm out} = I_{\nu,0}[1-P(h\nu)]+N_{K\alpha}\delta(h\nu-6.7 \mbox{
  keV})\cdot(6.7 \mbox{ keV}),
\end{equation}
where both $I_{\nu, \rm out}$ and $I_{\nu, 0}$ are measured in the
local frame of the disk. 

\begin{figure}[ht]
\caption{\label{fig:spectra_lines} Iron line absorption and emission
  features as measured by an observer at infinity, including all
  relativistic effects, for $\dot{m}=0.1$ and $i=30^\circ$.
  The curves show the
  ratio of the observed flux to that which would be observed without
  line physics included. The
  red curve shows only the absorption edge, the blue curve shows only
  the emission line, and the black curve shows the combination of
  absorption and emission. Above 10 keV, the sharp spectral features
  are due to Monte Carlo noise.}
\begin{center}
\includegraphics[width=0.9\linewidth]{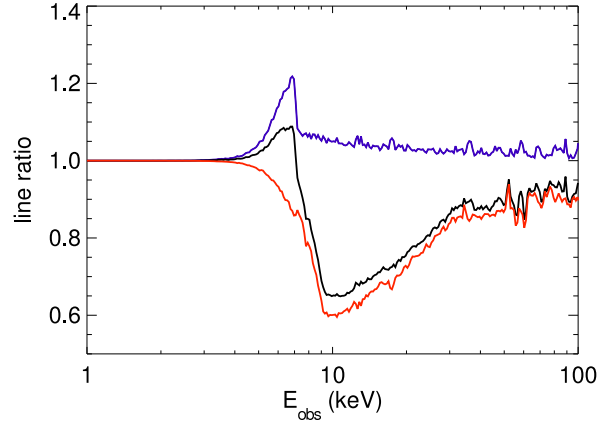}
\end{center}
\end{figure}

This outgoing photon packet then propagates towards the observer,
getting up-scattered by coronal electrons, possibly returning to the
disk, or captured by the black hole. The Fe K$\alpha$ emission line and
absorption edge are both broadened by relativistic effects as well as
the IC scattering. To show the magnitude and shape of these spectral
features, in Figure \ref{fig:spectra_lines} we plot (black curve) the
ratio of the total observed spectrum to what would be
observed if no absorption or emission were included in the
calculation. We also show the absorption and emission contributions
separately with the red and blue curves, respectively. 

When plotted as a ratio to the hypothetical no-line spectrum, the
absorption appears to dominate over the emission.  Two facts account
for this effect.  First, $f_{K\alpha}=0.5$ means that twice as many
photons get absorbed as emitted.  Second,
although in the rest-frame only photons with energy $> 8.8$~keV can be
absorbed and all line photons have energy exactly 6.7~keV, relativistic
broadening can shift part of the absorption feature downward
in energy and part of the emission line upward.  Where they overlap,
there is substantial cancellation.  Bluewards of the point where they
exactly cancel, the spectrum shows a sharp absorption feature that looks like
$1-P(E)$. Of course, the ratios in Figure \ref{fig:spectra_lines}
could never be directly observed, since they require knowledge of some
hypothetical spectrum that conveniently ignores fluorescent line
physics, but they do provide valuable insight into the physical
processes at work here.

In practice, we observe spectra like those shown in
Figure~\ref{fig:spectra_lum}
and then attempt to infer the shape of the emission line by fitting the
continuum with phenomenological models, an approach that can introduce serious
systematic errors (e.g., \citet{miller:07} and references therein).  The great
advantage of this global radiation
transport calculation is that {\it we can simultaneously fit the entire
spectrum with a single model based on physical parameters}---black hole
mass, spin, and accretion rate, Fe abundance, and observer inclination
angle, obviating the historical reliance on more phenomenological
models. Ultimately, we hope to apply our global radiation transport
techniques to a large body of MHD simulations, resulting in a
comprehensive suite of tabulated, self-consistent spectra that can be
incorporated into a standard X-ray spectra analysis package like {\tt
  XSPEC} \citep{arnaud:96}. 
As mentioned above, more detailed ionization physics
will be required before \pan can be used to fit real iron line data
with high precision.  Nonetheless, in the meantime we can still 
use the simulated spectra to gain important insights into the behavior
of the inner accretion flow. 

\begin{figure}[ht]
\caption{\label{fig:line_profile_inc} Iron line profile as a
  function of observer inclination, for $\dot{m}=0.01$. Only the
  emission contribution is shown, as the ratio to a hypothetical
  spectrum with no iron line physics included. The extended blue tail above 8
  keV is due to inverse-Compton scattering of the line photons in the
  corona.} 
\begin{center}
\includegraphics[width=0.9\linewidth]{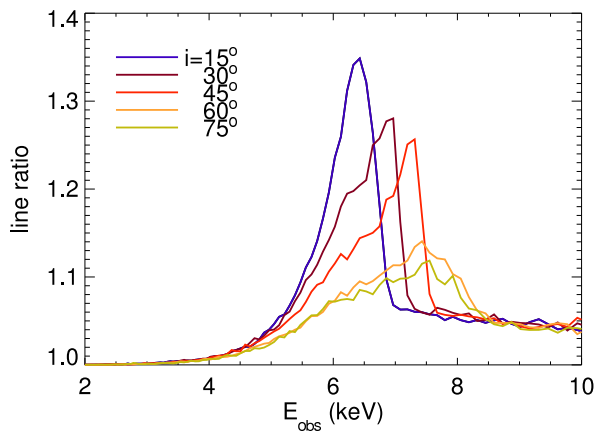}
\end{center}
\end{figure}

To focus on the relativistic effects of broad line physics,
in Figure \ref{fig:line_profile_inc} we plot the shape of the iron
line (ratio of ``emission only'' to ``no line physics'') for a range
of observer inclination angles for $\dot{m}=0.01$ (the thinner disk
highlights the relativistic effects). Again, we emphasize that these
line profiles cannot be directly observed, but only inferred after
fitting multiple spectral components such as the thermal peak, power
law, reflection hump, and smeared absorption edge
\citep{miller:04,miller:12}. 
%%JDS
More appropriately, these line profiles can be compared with
theoretical semi-analytic calculations well-known in the literature for
over two decades (e.g., \cite{laor:91}), which typically
assume planar circular orbits, and a K$\alpha$ surface brightness with a
power-law radial profile abruptly cut off at the ISCO.  Because these simple
models are the ones generally used in packages such as {\tt XSPEC}, they
define the range of line shapes to be fitted; contrasting them
with our profiles, based on more physical models for the hard X-ray
illumination, the disk mass profile, and the fluorescing material's
velocity, can indicate what might be achieved with future {\tt XSPEC}
models based more directly on disk physics.

One important feature that is not often discussed is the
high-energy tail clearly seen at all inclinations
\citep{petrucci:01}. This is due to IC scattering in the corona, 
the very process that generates the ionizing flux in the first
place. Since the coronal scattering is nearly isotropic, the amplitude
of the emission line above $\sim 8$ keV is independent of viewer
inclination. At the same time, the optical depth from the disk directly to a
distant observer increases with inclination angle, thereby reducing
the total number of line photons that {\it don't} get up-scattered, as can be
seen by comparing the integrated line flux below 8 keV.  The diminished
contrast between line and continuum associated with high inclination
angles may make it systematically more difficult to detect K$\alpha$
lines from those directions.
%%JDS
Again, the emission-only component of the high-energy tail would not
be observable directly, but would be combined with the dominant
power-law continuum. We present it here to focus on the underlying
scattering physics that produces the line.

\begin{figure}[ht]
\caption{\label{fig:line_profile_lum} Iron line profile as a
  function of luminosity, for observer inclination $i=30^\circ$. The
  red wings of the lines are remarkably similar, despite differences in
  $R_{\rm refl}$ and illumination profiles.}
\begin{center}
\includegraphics[width=0.9\linewidth]{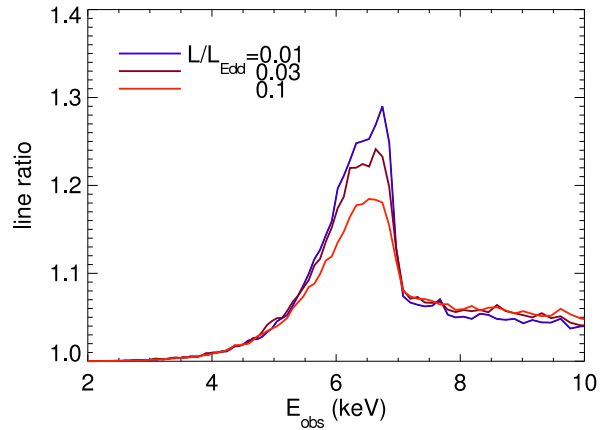}
\end{center}
\end{figure}

One of the most important potential applications of the iron line is
to use it as
a probe of where (or {\it if}) the disk truncates.  If there is a truncation
radius, and this radius can be quantitatively related to the ISCO, measurement
of the truncation radius could lead to a measurement of the black hole spin.
Indeed, many have attempted to measure spin {\it assuming} that such a sharp
truncation takes place exactly at the ISCO (a sampling of these efforts may
be found in \citet{martocchia:02,miller:02,duro:11,reis:11,reis:12,fabian:12}).
Even for the single spin value ($a/M=0$) simulated in ThinHR, we have
shown in Table~\ref{table:L_cor} and Figure~\ref{fig:ionizing_flux} that
the reflection edge of the disk can be adjusted by modifying 
$\dot{m}$.  Therefore we might reasonably expect very different line
profiles for $\dot{m}=0.01$, 0.03, and 0.1, corresponding to average
reflection edge radii $R_{\rm refl}/M=6.1$, 4.4, and 2.1,
respectively.  Gravitational
redshift is especially strong in the plunging region, so sizable contrasts
in the red portion of the profile might be expected.  A
sample of line profiles is shown in Figure \ref{fig:line_profile_lum}
for a range of $\dot{m}=0.01-0.1$ (for $\dot{m}>0.1$, the finite
thickness of the disk begins to distort the shape of the line,
confusing the dependence on $R_{\rm refl}$), holding the observer
inclination constant at $i=30^\circ$. Remarkably, these lines show
very little variation with $R_{\rm refl}$, especially in the red
wing below 6 keV. Only the overall normalization is different, with
somewhat weaker lines when $\dot{m}$ is larger, due to the the softer
ionizing spectrum around 9 keV. 

%The only differences evident are the height of the peak and the
%amplitude of the IC tail. Both of these features may be attributed to the
%different effective geometries, as the higher luminosity cases have a
%greater scale height for the disk and corona, somewhat increasing the
%optical depth through which a line photon must pass before reaching
%the observer. In the limit
%where the observer inclination angle is the same as the disk opening
%angle, the optical depth through a sandwich-type corona becomes
%infinite, and it becomes harder for line photons to escape directly to
%the observer without scattering.
%This finite-opening-angle effect also explains the different shape of the line for
%$\dot{m}=1.0$.  With a photospheric aspect ratio $H_{\rm phot}/r\approx 0.5$,
%its configuration can hardly
%be called a thin disk. In that case, an observer at $30^\circ$ would
%actually be sampling a range of effective inclinations between
%$0^\circ$ and $60^\circ$, and thus the total integrated line looks
%like a combination of these different viewing angles. 

This lack of sensitivity to the inner disk location is due
to the fact that gas inside of the ISCO is already plunging rapidly
towards the horizon. Most of the line photons produced
in the plunging region get beamed into the black hole, never reaching a
distant observer. Similar results were seen for thermal emission from
the plunging region in \citet{zhu:12}. As a test of this effect, we
compared the total flux that eventually reached infinity with that
which was captured by the horizon as a function of the radius of the
initial seed photons. For $\dot{m}=0.1$, $40\%$ of the seed
photons emitted from $r=5M$ got captured by the horizon, a fraction that
climbs to $95\%$ at $r=3M$.
As can be seen in Figure~\ref{fig:spectra_r}, $\lesssim 1\%$ of the total
flux around 6 keV comes from inside of $6M$, regardless of where the disk
reflection edge is.
In future work making use of \harm simulations for
a range of spin parameters, we will explore whether the emergent profiles
have sufficiently strong dependence on spin that this diagnostic can
be successfully used. 
%%JDS
It will also be crucial to better understand how to deconvolve the
emission line profile from the other features of the observed
spectrum.

\section{BULK COMPTONIZATION}\label{section:bulk}

In addition to the thermal IC processes described above, the corona
can also transfer energy to the seed photons through ``bulk
Comptonization'' when the fluid velocity of the corona is large
relative to the disk. Some authors have used this process to explain
the hard tail seen in some thermal state observations \citep{zhu:12}, or
the steep power-law state when the bulk flow is
convergent \citep{titarchuk:02,turolla:02} or turbulent
\citep{socrates:04,socrates:10}. To quantify this effect in the \harm
simulations, we simply set the electron temperature everywhere in the corona to
zero while maintaining the turbulent motion above the disk and the
convergent flow in the plunging region. 

\begin{figure}[ht]
\caption{\label{fig:bulk_compton} Net Compton power $dL/d(\log r)$ in the
  corona when $T_e=0$, normalized to the total seed luminosity from
  the thermal disk. Note the unusual labeling of the $y$-axis, with
  logarithmic scaling above and below $0=\pm 10^{-4}$.}
\begin{center}
\includegraphics[width=0.9\linewidth]{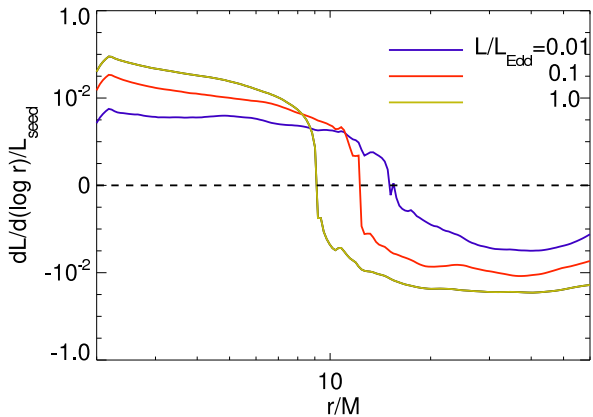}
\end{center}
\end{figure}

As before, we calculate the total Compton power in each fluid element
by subtracting the energy in the incoming photon packet (as measured
at infinity) from that of the outgoing ray. For an electron at rest,
the photon will always transfer energy to the electron, giving
negative IC power (hence the distinction between Compton scattering
and {\it inverse} Compton scattering). In Figure
\ref{fig:bulk_compton} we plot the 
bulk coronal power in terms of $dL/d(\log r)$, normalized by the total disk
luminosity, for a range of accretion states. The $y$-axis is
logarithmic and signed, so we set $10^{-4}=0=-10^{-4}$ for improved
visualization. Where $dL/d(\log r)>0$, the bulk velocity
of the gas transfers energy to the radiation field. Where $dL/d(\log r)<0$,
the typical energy of a seed photon is greater than the bulk kinetic
energy of the coronal electrons, and the radiation field loses
energy to the corona.

Three conclusions may be drawn from Figure \ref{fig:bulk_compton}: (1)
bulk Comptonization plays a very minor role in the overall energetics of thin
accretion disks; (2) bulk Comptonization is most significant for high
accretion rates; and (3) for all values of $\dot{m}$, $dL/dr>0$ in the
inner regions and $dL/dr<0$ in the outer disk. The explanations of
these effects are straightforward: (1) turbulent velocities in the
corona are simply not very large. In the plunging region, there are
either not enough seeds (low $\dot{m}$) or the optical depth is high
(large $\dot{m}$), so the seeds are advected into the black hole without being
able to sample a wide range of converging velocities. Not
surprisingly, the observed spectra for the bulk Comptonization runs
are nearly indistinguishable from pure thermal disks. (2) Large
$\dot{m}$ corresponds to large $H_{\rm phot}/r$, and we find
that the turbulent velocities generally increase with scale height
above the disk, so higher luminosity systems are sampling more
turbulent regions of the corona. (3) For disks with constant $H_{\rm phot}/r$,
turbulent velocity should scale like the orbital velocity $v_{\rm
  bulk} \sim v_{\rm orb} \sim r^{-1/2}$, so the turbulent kinetic
energy scales like $r^{-1}$. The seed photon energy, on the other hand,
scales like $r^{-3/4}$ in the outer disk, and actually begins to
decrease in the inner region as the disk becomes optically thin. 

\begin{figure}[ht]
\caption{\label{fig:phot_bulk} Seed photon energy (dashed curves) and
  specific turbulent kinetic energy of the fluid in the corona (solid
  curves) for $L/L_{\rm Edd} = 0.01, 0.1, 1.0$.}
\begin{center}
\includegraphics[width=0.9\linewidth]{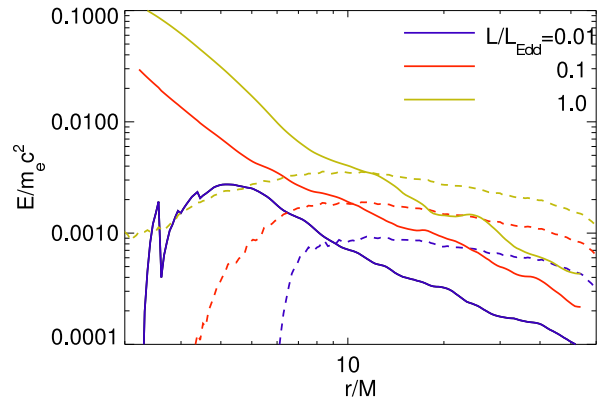}
\end{center}
\end{figure}

In Figure \ref{fig:phot_bulk} we show the average kinetic energy in the
corona as a function of radius (solid curves), along with the seed photon
energy (dashed curves). Here, the specific kinetic energy is defined
as $1/2\, \sigma_v^2(r)$, where $\sigma_v(r)$ is the variance of the
3-velocity $u^i$, sampled over all $\phi$, $t$, and optical depth
$\tau=0.1-1$ at each radius. Outside of $r\approx 10M$, the photon energy
is higher, and thus transfers energy into the corona, giving
$dL/dr<0$. Note that, for $\dot{m}=0.01$, the nearly laminar flow at
the midplane is considered part of the corona, not the disk, thus
explaining the turnover in turbulent kinetic energy inside of $\sim
4M$. 

\section{X-RAY VARIABILITY}\label{section:variability}

\begin{figure}[ht]
\caption{\label{fig:light_curves} X-ray light curves for $\dot{m}=0.1$
  and viewer inclination of $i=60^\circ$, taken from 
  simulation snapshots sampled every $500M$ in time during the period
  of inflow equilibrium. In each energy band, the light curve is
  normalized to the
  mean flux in that band, and divided by the linear trend of the
  bolometric flux.}
\begin{center}
\includegraphics[width=0.9\linewidth]{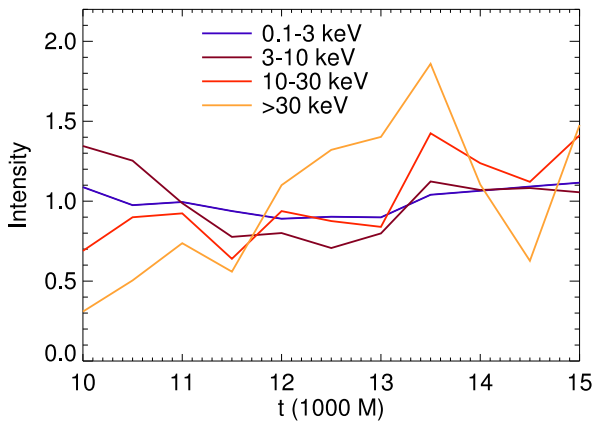}
\end{center}
\end{figure}

Up to this point, all the discussion in this paper has focused on
steady-state behavior of the simulated spectra. As mentioned in
section \ref{section:harm}, the results are based on snapshots of the
ThinHR simulation, spaced every $500M$ between $10000M$ and $15000M$,
roughly the period of inflow equilibrium. With these 11 snapshots, we
are also able to carry out some very coarse timing
analysis. Figure \ref{fig:light_curves} shows simulated light curves in four
different energy bands, for $\dot{m}=0.1$ and observer
inclination $i=60^\circ$. Over the period shown, the bolometric
luminosity of the simulation changes by about $20\%$, quite typical of
global MHD simulations. To focus on the intrinsic variability, we have
normalized all light curves by a single linear trend over this
period. In Figure \ref{fig:light_curves}, each individual light curve
has also been normalized by its mean value, to show the relative
amplitude of fluctuations. 

In Table \ref{table:xray_rms} we list the RMS variation in the flux in
different energy bands for different values of $\dot{m}$, again normalized by the
linear trend in the bolometric luminosity. At low energies,
corresponding to the thermal peak, we see a clear
anti-correlation between accretion rate and variability, due to
the fact that at low $\dot{m}$, the inner disk is moving in and out,
changing the thermal seed flux. The low variability of the case with
$\dot{m}=1.0$ further strengthens its classification in the thermal
state \citep{remillard:06}. 

Our findings show an opposite trend in variability with $\dot{m}$ than found
in \citet{noble:09b}. They measured the bolometric flux from dissipation
in the coronae only. Since they neglected the seed flux
from the thick disk and ignored all IC scattering/heating physics in the
corona, their only source of variability was the
intrinsic dynamic variability of coronal turbulence and
dissipation. Thus their trend of increasing variability with increasing
$\dot{m}$ suggests larger intrinsic fluctuations at higher altitudes
above the disk, consistent with increased
turbulent velocities, as shown above in Figure \ref{fig:phot_bulk}.

%new references and distinction b/t hard, soft states
At all $\dot m$, there is a clear increase
in variability with photon energy, as seen in observations of the
thermal dominant and steep power-law states of some black holes
\citep{cui:99, churazov:01, gierlinski:05}.
%This could result if the high-energy flux comes from a power law
%with variable index, where the power-law tail ``pivots'' around the thermal
%peak like a see-saw \citep{cabanac:09}. 
When there is evidence for a
thermal disk in the hard state, this same energy-RMS relation is seen
\citep{wilkinson:09}. On the other hand, for hard states with no clear
evidence for a disk, the RMS appears to be constant or even decrease
with energy \citep{nowak:99a,gierlinski:05}. 

One possible explanation for the different RMS-energy scaling found in
different states is that for softer states, the seed photons generally
have higher energies and thus can get scattered up to $\gtrsim 50$ keV
after only a few scatterings, which is possible to do within a single
coronal hot spot. For the hard state, the lower-energy seed photons
will typically need more
scatterings to reach the same energy, and thus sample a larger volume
of the corona, averaging over many hot spots. This explanation is
consistent with the RMS variability seen in AGN \citep{vaughan:03},
which increases with energy at low energy ($\lesssim 1$ keV; few
scatters in a single hot spot) and then decreases at high energy
($\gtrsim 1$ keV; many scatters throughout entire corona). 
Further supporting this suggestion is the increase in RMS with AGN
flux \citep{vaughan:03}: as the source brightens, the seed photon
energy increases, and can more easily get boosted to keV energies
within localized hot spots. 

It is apparent from Figure \ref{fig:light_curves} that the light curves
in the different bands are not tightly correlated, as would be expected if
the variability were strictly due to global coronal properties like the
Compton $y$-parameter.  That lack of correlation indicates that the
dependence of variability on photon energy is due to fluctuations that
are independent in regions of different temperature, as well as stronger
in regions of higher temperature.  Such a situation is a natural result
of variability in local heating. 

\begin{table}[ht]
\caption{\label{table:xray_rms} RMS variability in different energy
  bands as a function of luminosity, for an observer at $i=60^\circ$. 
  To remove the secular trend, the light curves in each energy band have 
  been normalized by dividing out a linear fit to the bolometric
  luminosity.}
\begin{center}
\begin{tabular}{lccccc}
\hline
\hline
$L/L_{\rm Edd}$ & 0.1-3 keV (\%) & 3-10 keV (\%) & 10-30 keV (\%) & $> 30$
keV  (\%) \\ 
\hline
0.01 & 15  & 24 & 35 & 54 \\
0.03 & 14  & 26 & 40 & 56 \\
0.1  & 12  & 24 & 32 & 58 \\
0.3  & 8.4 & 21 & 26 & 54 \\
1.0  & 6.8 & 19 & 28 & 51 \\
\end{tabular}
\end{center}
\end{table}

\begin{figure}[ht]
\caption{\label{fig:rms_inc} Fractional RMS amplitude (\%) for
  azimuthal variations in the observed flux for $\dot{m}=0.1$, as a
  function of viewer inclination and photon 
  energy. The color code is the same as in Figure
  \ref{fig:light_curves}. The error bars correspond to the 1$\sigma$
  range of RMS values calculated for each snapshot in time.}
\begin{center}
\includegraphics[width=0.9\linewidth]{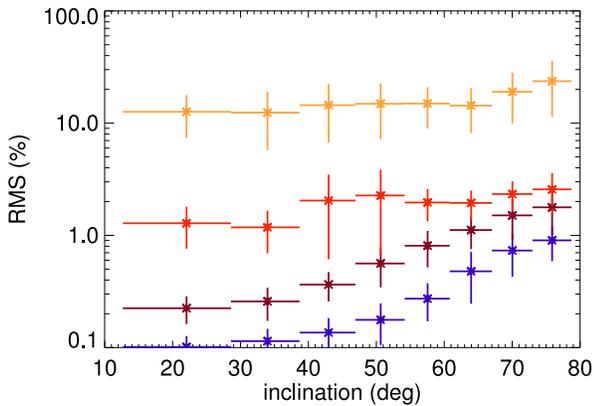}
\end{center}
\end{figure}

To get a better estimate of the local coronal effects on the light
curves, we have also calculated phase-dependent light curves by
calculating the flux seen by observers at different azimuths. The
variability as a function of observer $\phi$ is a proxy for continuum
fluctuations at high frequencies, comparable to the orbital frequency,
where variability is often
quite strong in the hard and steep power-law states \citep{remillard:06}. To estimate
the amplitude of these modulations, we construct many short light
curves, one for each snapshot and inclination angle, and calculate the
fractional RMS amplitude for each light curve. We then average over
all these snapshots, plotting the mean RMS (along with $1\sigma$ error
bars) in Figure \ref{fig:rms_inc} as a
function of observer inclination for a range of energy bands. 
This procedure is roughly
equivalent to measuring the variance in the observed light curves over
a narrow frequency band corresponding to the orbital frequencies of
the parts of the disk that contribute the most power.
In order
to resolve the high-energy fluctuations, we use a particularly large
number of Monte Carlo photon packets, roughly $10^9$ rays per snapshot.

The fractional RMS amplitude
rises steadily with inclination, consistent with a non-axisymmetric
source because relativistic beaming in the orbital direction is
greatest for edge-on observers.  In contrast, the variability created
by global axi-symmetric modes is greatest for face-on
observers \citep{schnittman:06a}. Similarly, \citet{noble:09b} did
not find a strong correlation of
fractional variability with inclination, likely due to the fact that
they used an optically thin ray-tracing procedure, neglecting
any coronal scattering. 

The fact that the RMS amplitude
increases with energy---as seen in observations
\citep{remillard:06}---suggests that the variability is coming from
the corona and not from 
the disk. Any fluctuations in the seed photons would be
smoothed out when propagating through a uniform corona, just as
pulses from a lighthouse are dispersed in fog, and would give lower
variability at high energy because larger numbers of scatterings
are required for the seeds to reach high energy \citep{schnittman:05}.
Combined, these results are highly suggestive of a coronal hot-spot model for
high-frequency X-ray variability in black hole binaries.

\section{COMPARISON WITH CLASSICAL DISK THEORY}\label{section:disk_theory}

The spectral states of galactic black hole binaries are roughly correlated
with their bolometric luminosities in the sense that low luminosity
states generally have hard spectra, while higher luminosities
permit a broader range of spectral states but exhibit a preference
for softer states \citep{fender:04,remillard:06}. Our model is able to
reproduce this observed correlation, yet does so in a fashion that
differs in several respects from classical disk theory.  In our model,
increased accretion rate leads to a proportionately larger surface
density, but leaves the scale height $H_{\rm dens}$ unchanged.  The systematic shift
in spectral shape with accretion rate is due to a change in how the
corona and the thermal disk share the dissipation: as the accretion
rate increases, the photosphere moves to larger multiples of $H_{\rm dens}$, so
that more of the dissipation occurs within the thermal disk.  By
contrast, in classical disk theory, both the surface density $\Sigma$ and $H_{\rm dens}$ are functions
of accretion rate \citep{shakura:73}.  As we will show in a moment, most of our
parameter space lies in the radiation-dominated regime, in which
%%%JHK: a couple of clarifying words
classical disk theory predicts
$\Sigma \propto \dot{m}^{-1}$ and $H_{\rm dens} \propto \dot m$.  Moreover,
for the accretion rates
we consider, classical disk theory assumes that ${\it all}$ the dissipation
takes place inside the disk, leaving no room for a corona at all, and
identifies the photosphere precisely with $H_{\rm dens}$.  These contrasts raise
two questions: Is there any $\dot m$ at which our model and classical
theory overlap?  And should the trends with $\dot m$ predicted by the
single-simulation model presented here be expected to carry over to
global MHD simulations with different values of $H_{\rm dens}$, corresponding
to different values of $\dot m$?

As shown in Figure~\ref{fig:spectra_r}, most of the light is produced
in the range $r \simeq 6$--$30M$ for all the accretion rates we studied.
Our account of the predictions of conventional disk theory therefore centers
on that range.  According to this theory, radiation pressure exceeds
gas pressure inside disks when the accretion rate is greater than
\begin{equation}\label{eq:mdotrg}
\dot m_{rg} \simeq 0.02 \alpha_{SS}^{-1/8} (M/M_{\odot})^{-1/8} 
            (r/10M)^{21/16}(R_R/0.2)^{-9/8} R_T^{1/8} R_z^{5/8},
\end{equation}
where $\alpha_{SS}$ is the usual ratio of vertically-integrated 
fluid-frame stress to vertically-integrated pressure, $R_R \leq 1$ is a
function of radius that adjusts the vertically-integrated
fluid-frame dissipation rate for both the net angular momentum flux through
the disk and relativistic corrections, $R_T \simeq R_R$ is
a similar correction factor applied to the vertically-integrated
fluid-frame stress, and $R_z$ (usually slightly greater
than unity) introduces relativistic corrections into the vertical component
of gravity (notation as in Krolik 1999).  Because $R_R$ increases outward
in this range of radii, $\dot m_{rg}$ rises only gradually with radius.
Thus, almost all the span of accretion rates we consider falls into the
radiation-dominated regime.

Conventional analytic disk theory estimates disk thickness in
the radiation-dominated limit by supposing that radiation provides all
the disk's support against the vertical component of gravity and that
all dissipated energy is conveyed outward by radiation flux
\citep{shakura:73,novikov:73}.  This
pair of assumptions combined with the condition of hydrostatic
equilibrium leads to the conclusion that all of the dissipation within
the disk must be accomplished within a distance
\begin{equation}\label{eq:radheight}
H_{\rm rad} = (3/2)(\dot m/\eta) \left(\frac{G M}{c^2}\right) \frac{R_R}{R_z}
\end{equation}
of the disk midplane.
Thus, for radiation-dominated disks, scale height and
accretion rate are directly proportional, and one can be used as a proxy
for the other.  Conventional theory also assumes that
the density is constant for $|z| \leq H_{\rm rad}$ and zero outside
$H_{\rm rad}$
\citep{shakura:73}.  Because this theory has no explicit place for
coronae, it is unclear what luminosity it predicts for them, but
presumably any corona begins outside $H_{\rm rad}$.

Detailed stratified shearing box simulations of radiation-dominated disk
segments \citep{hirose:09,blaes:11} have shown
that $H_{\rm rad}$ does give an order of magnitude estimator of the vertical
scale height of such disks, but, unsurprisingly, the density distribution
is crudely exponential.  Because both gas and magnetic pressure can contribute to
vertical support and some vertical energy transport is by radiation advection
rather than diffusive flux, these simulations also find that the dissipation
profile is more extended than indicated by conventional analytic theory.
In particular, only about half the dissipation takes place within a distance
$H_{\rm rad}$ of the midplane, and 90\% of the dissipation is accomplished inside
$\simeq 2H_{\rm rad}$.  The photosphere is generally found at
3--$4H_{\rm rad}$.  These simulations
were conducted with surface densities $\sim 10^4$--$10^5$~g~cm$^{-2}$, a
range relevant to the larger radius and smaller accretion rate portion
of our parameter space, so the specific numbers quoted might require
adjustment for smaller radii and larger accretion rates.

The parallel set of facts about our simulated disk is that its density
scale height is always $H_{\rm dens} \simeq 0.06r$, but the height of its photosphere is a
multiple of $H_{\rm dens}$ that increases with $\dot m$, ranging from $\simeq 2$ to
$\simeq 9$ as $\dot m$ increases from 0.01 to 1.0 (see Table~\ref{table:L_cor}).
This is, of course, why the corona's share of the luminosity $f_{\rm
  cor}$ decreases with
increasing $\dot m$ in our model, as explained above in Section \ref{section:cor_temp}.

Having placed the two pictures side-by-side, we can now locate the parameters
for which they resemble each other.  Because $H_{\rm rad} \propto R_R$ and $R_R$ scales
with $r$ somewhat more slowly than linearly for Schwarzschild spacetimes between
$r=10M$ and $r=30M$, while the simulation has $H_{\rm dens} \propto r$, it is possible to match
the simulation photosphere to the photosphere predicted by classical disk theory
across this range of radii. We find approximate agreement when $\dot m \simeq 0.2$.

We now turn to the question of what to expect from simulations with different
$H_{\rm dens}(r)$, representing other values of $\dot m$.
%%%
In the radiation-dominated regime, classical theory predicts $H_{\rm dens} \propto \dot m$.  The spectral predictions
made here show the spectrum becoming more thermal as $\dot m$ increases because
increasing $\dot m$ leads to a larger share of the total dissipation taking
place inside the photosphere.  Thus, with regard to the proportion of the
light in the thermal part of the spectrum, the central issue is whether, as $\dot m$
and therefore $H_{\rm dens}$ increase, $H_{\rm phot}/H_{\rm dens}$
rises more or less rapidly than $H_{\rm diss}/H_{\rm dens}$.
%%%
If, as in our single-simulation model, the former increases more rapidly with
$\dot m$ than the latter, our predictions on this issue will be (at least
qualitatively) vindicated; if not, they will need revision.
%%%

%%%
The most important scaling issue for the corona is likewise the fraction
of the dissipation it receives.  At the very crudest level, a global model
for the corona would yield a luminosity $\propto f_{\rm cor}\dot{m}$, where
the fraction $f_{\rm cor}$ of the dissipation going into the corona can
be a function of $\dot{m}$ (e.g., in our picture, a decreasing one
with $f_{\rm cor}\sim \dot{m}^{-1/3}$).
Our predictions for how the hard X-ray luminosity scales with $\dot m$ therefore
stand or fall on the relative scaling of $H_{\rm phot}$ and $H_{\rm diss}$
just as our predictions for the thermal portion do.  

The shape of the coronal spectrum depends in a general way on its mean temperature
and optical depth, but the {\it distribution} of dissipation and density also
play a significant role.
%%%JHK: A few bridging words
To estimate how they might change in new simulations with a dependence of
$H_{\rm dens}$ on $\dot{m}$ mimicking the one predicted for radiation-dominated disks,
%%%
we use the scaling argument of equation (\ref{eqn:T_LU}) because it is
actually quite general.  Although it assumes that the corona's
total vertical optical depth is always unity, the way the coronal temperature
falls as the photosphere is approached suggests that this is a reasonably good
approximation, and one unlikely to be much affected by changes in disk parameters.
Its detailed development, leading to equation (\ref{eqn:Tscale}), however,
depends on an overall density scale that is $\propto \dot m$, and this can
change with parameters.  Nonetheless,
%%%JHK: A few more clarifying words
for any relation between density and $\dot{m}$,
%%%
within the corona one still finds that
$\gamma^2\beta^2 \propto ({\mathcal L}_{\rm phot}/U_{\rm ph}) \tau^{-1}$.  The
question is how the ratio of dissipation to photon density at the photosphere
depends on $\dot{m}$.  This ratio is, of course, simply $f_{\rm cor}$.

As we emphasized earlier, the spectral shape is also influenced
by the spatial distribution of heating, in the sense that greater inhomogeneity
tends to yield harder spectra.  The range of temperatures seen in
the corona is primarily determined by the range in the local heating rate
per unit mass.  In our simulation, ${\mathcal L}/\rho$ increases outward through
the corona.  It seems plausible that this trend would be reproduced
in simulations with different scale heights because it largely reflects
magnetic buoyancy, an effect that should be universal in magnetized accretion
flows.
%%%JHK: Another small wording fix
However, the robustness of this trend can be checked
%%%
when additional simulations with other scale heights are performed.  Similarly,
we know of no reason why the statistics of inhomogeneity should change, but future
simulations will reveal whether this is so.
%%%

\section{DISCUSSION}\label{section:discussion}
Using the new radiation transport code \pand, we have analyzed
data from a high-resolution, 3D MHD simulation of accretion onto a
Schwarzschild black hole. Because the MHD code \harm is energy
conserving, we are able to employ a cooling function that tracks the
local dissipation of energy throughout the simulation volume. By
combining the results from \pan and \harmd, we have for the first time
been able to produce a global, self-consistent solution for the
radiation field around an accreting black hole, and predicted---on the
basis of real physics---the coronal luminosity.
The major results from this work can be summarized as follows:

\begin{enumerate}
\item We have shown---for the first time---that
%%%JHK: Tightened the wording
  MHD turbulence in an accretion disk can lead to dissipation
  outside the disk's photosphere strong enough to power hard X-ray
  emission comparable in luminosity to the disk's thermal luminosity.
  This is a result long-expected, but never before demonstrated
  directly. 
\item For different values of the Eddington-normalized accretion rate
  $\dot{m}$, the location of the photosphere changes, in turn varying
  the fraction of radiative power in the disk (thermal) and the corona
  (inverse Compton). 
  The coronal temperature ranges from about 10 keV near the disk
  surface up to $\sim 100$--300~keV in the upper, low-density
  corona.
%JS%
  At a fixed optical depth above the photosphere, we find the corona
  temperature increases slowly with decreasing accretion rate.
%  Independent of black hole mass, the corona temperature
%  scales inversely with accretion rate:
%  $T_e \sim \dot{m}^{-1}$ at temperatures well below $m_e c^2/k_B$ and
%  $T_e \sim \dot{m}^{-1/2}$ in the relativistic regime.

\item By varying $\dot{m}$ from 0.01 to 1, we naturally reproduce
  X-ray spectra consistent with those observed in the hard, steep
  power-law, and thermal states of galactic black hole binaries. The
  spectra are characterized by a thermal
  peak around 1 keV and a high-energy power-law tail extending above
  100 keV.
%%%
  Although the fraction of the total heat released by the
  corona is never greater than $\simeq 40\%$, the highly inhomogeneous
  dissipation predicted by the MHD simulation provides the local concentration
  of heating necessary to create output spectra as hard as observed.
  %JS%
%  previous Comptonization models found it necessary to invent geometries
%  to satisfy this criterion \citep{haardt:94,stern:95,zdziarski:96,poutanen:97}.
%%%
  In most cases there is evidence for a Compton reflection
  hump between 30 keV and 100 keV. 
\item The Fe K$\alpha$ illumination profile of the disk follows the classical
  $r^{-3}$ scaling at large radius, then flattens to $r^{-3/2}$ in the
  inner disk. At lower values of $\dot{m}$, the disk begins to
  disappear inside the ISCO, and the line production is
  correspondingly reduced. 
  The iron line profiles consist of both an absorption edge above 8.8
  keV, and a broad emission line around 6.7 keV with a strong red-shifted
  tail. The shape of the line is dependent on observer inclination,
  and in all cases has a significant tail above 8 keV due to
  up-scattering of the line photons in the corona.

  The broad iron line profile appears to be only weakly dependent on
  the location of the disk inner edge, as most
  photons generated in the plunging region never reach the
  observer. Thus iron lines may in fact be better at
  measuring the location of the ISCO than the disk's reflection edge. 
\item Bulk Comptonization plays a very minor role in the photon
  energetics, typically $\lesssim 1\%$ of the total seed luminosity. In
  the outer region of the disk $(r\gtrsim 10M)$, the thermal seed
  photons carry more energy than the bulk kinetic energy in
  the coronal electrons.
\item We have carried out some initial timing analysis of the
  simulated X-ray spectra, and find a number of trends that are
  consistent with observations: the
  fractional RMS amplitude increases with decreasing luminosity, and
  for all accretion rates, the RMS amplitude increases with photon
  energy. On short time scales, the variability increases with
  observer inclination and photon energy, as expected for a coronal
  hot-spot model of X-ray variability. 
\end{enumerate}

Although the progress made to date has been significant, this work
is just the tip of the iceberg. We are currently in the
process of analyzing new \harm simulations, carried out with
resolution comparable to that of ThinHR, for a wide range of black hole spin
parameters. This will allow us to explore both potential dependence
of the disk/corona continuum spectrum on spin and  greatly improve our understanding of
the iron line as a probe of black hole spin, disk dynamics near the ISCO, and
the nature of the plunging region.  New simulations with high spin will
also allow us to probe the properties of the relativistic
jet, frequently seen in observations of the hard state, and most
clearly present in simulations of spinning black holes
\citep{mckinney:04,hawley:06}. 

In other future work, simulations of disks with different $H_{\rm dens}(r)$ profiles
will expand the applicability of our models to a wider
range of X-ray states and accretion rates. We will explore what parameters other than
$\dot{m}$ and $H_{\rm dens}(r)$ determine the state of a given black hole. Improving the physics
of the fluorescent line, including ionization balance and more detailed
excitation cross sections \citep{garcia:11a, garcia:11b} will make
predictions of its strength and profile more reliable.  Including
the energy lost to photoionization and Compton recoil in the disk surface will also
permit a better treatment of hard X-ray reprocessing and that energy's
reemergence in disk continuum, processes relevant to AGN.

We also plan on extending our preliminary variability analysis to the
entire set of simulation data within ThinHR's statistically steady
epoch (over 5000 snapshots in time), allowing for a more detailed study
of high-frequency fluctuations and the possible identification of
quasi-periodic oscillations (QPOs), sometimes seen in galactic black
holes in the steep power-law state \citep{remillard:06}. In
addition to QPOs, we should be able to characterize time lags between
different energy bands as a function of frequency, and compare with a
large body of observational results (e.g.,
\citet{nowak:99}). These lags appear to scale like the light-crossing
time for fluctuations in the thermal seed flux to propagate
through the corona, and thus could be a powerful probe of the coronal
geometry \citep{uttley:11}. We can also investigate the effects of
finite light-travel time through the simulation volume, which was
found to suppress variability in \citet{noble:09b}. 

\pan was originally developed to study X-ray polarization
\citep{schnittman:09,schnittman:10}, so that information comes along for
free with all the calculations described in this paper.  As techniques
for high-sensitivity X-ray polarimetry continue to improve \citep{black:10},
polarization predictions will become observationally testable; it will
be interesting to compare predictions made from MHD simulations with the toy
coronal models presented in \citet{schnittman:10}. Because our analysis includes
broad-band spectra, line profiles, timing, and polarization
information in a single self-consistent calculation, it is the ideal
tool for integrating these complementary techniques for
measuring black hole spin and probing the physical properties of the accretion
flow. 

Despite the remarkable progress we have made in bridging the gap
between simulation and observation, there still exist numerous
challenges and caveats. The state-of-the-art MHD simulations still do
not include adequate thermodynamics or internal radiation transport
coupled directly with the fluid dynamics. While much
progress has been made in shearing-box simulations \citep{hirose:06,hirose:09},
there remain serious conceptual and computational obstacles to
incorporating these advances into global simulations. 
In the short-run, $H(r)$ profiles designed to mimic the mean effects of
radiation pressure support can help, but in the long-run it will be
necessary to include radiation forces explicitly in order to explore
what effect they have on both temporal and spatial fluctuations in the
structure of the corona. 

The ray-tracing
tools described here are also lacking in certain regards. Because of
the photon-packet methodology used in \pand, we have been forced to use
energy-independent scattering cross sections, which certainly breaks
down at high photon energy. Similarly, the angular distribution of
Fe~K$\alpha$ lines emerging from the disk is assumed to be identical
with the angular distribution of photons in the same packet reflected
by electron scattering as given by \citet{chandra:60}. These shortcomings can
be improved with relatively little effort, but at a cost to
computational efficiency. 

Without a doubt, the most important next step is the direct comparison
of our \pan spectra with real X-ray data. To this end, we are
fortunately blessed with a mass of archival data from {\it RXTE}, {\it
  Chandra}, {\it XMM-Newton}, and {\it Suzaku} with which to test our
spectral models and improve upon earlier phenomenological analysis
methods. 

\acknowledgements{
We would like to thank C.\ Done, A.\ Fabian, T.\ Kallman, and C.\
Reynolds for helpful discussions. 
This work was partially supported by NSF grants AST-0507455 and
AST-0908336 (JHK) and AST-1028087 (SCN).
The ThinHR simulation was carried out on the Teragrid Ranger
system at the Texas Advance Computing Center, which is supported in part 
by the National Science Foundation.}

\newpage

\end{document}